\documentclass[dvips]{arxstspdf}
\usepackage{graphics}
\usepackage{flushend}
\usepackage{stfloats}

\volume{24}
\issue{3}
\pubyear{2009}
\firstpage{361}
\lastpage{385}
\doi{10.1214/08-STS276}
\makeatletter
\setattribute{abstract}    {skip} {20\p@}
\setattribute{frontmatter} {skip} {8\p@ plus 3\p@ minus 3\p@}
\setattribute{frontmatter} {cmd}  {}
\setattribute{abstract}   {width}  {36.5pc}
\makeatother

\begin{document}
\begin{frontmatter}

\title{A Conversation with Leo Goodman}
\runtitle{A Conversation with Leo Goodman}

\begin{aug}
\author{\fnms{Mark P.} \snm{Becker}\ead[label=e1]{mbecker@gsu.edu}\corref{}}
\runauthor{M. P. Becker}

\affiliation{Georgia State University}

\address{Mark P. Becker is President,
Georgia State University, Office of the President,
P.O. Box 3999, Atlanta, Georgia 30302-3999,
USA \printead{e1}.}
\end{aug}

\begin{abstract}
Leo A. Goodman was born on August 7, 1928 in New York City. He
received his A.B. degree, summa cum laude, in 1948 from Syracuse
University, majoring in mathematics and sociology. He went on to pursue
graduate studies in mathematics, with an emphasis on mathematical
statistics, in the Mathematics Department at Princeton University, and
in 1950 he was awarded the M.A. and Ph.D. degrees. His statistics
professors at Princeton were the late Sam Wilks and John Tukey. Goodman
then began his academic career as a statistician, and also as a
statistician bridging sociology and statistics, with an appointment in
1950 as assistant professor in the Statistics Department and the
Sociology Department at the University of Chicago, where he remained,
except for various leaves, until 1987. He was promoted to associate
professor in 1953, and to professor in 1955. Goodman was at Cambridge
University in 1953--1954 and 1959--1960 as visiting professor at Clare College
and in the Statistical Laboratory. And he spent 1960--1961 as a visiting
professor of mathematical statistics and sociology at Columbia
University. He was also a research associate in the University of
Chicago Population Research Center from 1967 to 1987. In 1970 he was
appointed the Charles L. Hutchinson Distinguished Service Professor at
the University of Chicago, a title that he held until 1987. He spent
1984--1985 at the Center for Advanced Study in the Behavioral Sciences in
Stanford. In 1987 he was appointed the Class of 1938 Professor at the
University of California, Berkeley, in the Sociology Department and the
Statistics Department. Goodman's numerous honors include honorary D.Sc.
degrees from the University of Michigan and Syracuse University, and
membership in the National Academy of Sciences, the American Academy of
Arts and Sciences, and the American Philosophical Society. He has also
received numerous awards: From the American Statistical Association, the
Samuel S. Wilks Medal; from the Committee of Presidents of Statistical
Societies, the R. A. Fisher Lectureship; and from the Institute of
Mathematical Statistics, the Henry L. Reitz Lectureship; also, from the
American Sociological Association, the Samuel A. Stouffer Methodology
Award and the Career of Distinguished Scholarship Award; and from the
American Sociological Association Methodology Section, the inaugural
Otis Dudley Duncan Lectureship. Earlier he had received a Special
Creativity Award from the National Science Foundation, and fellowships
from the Guggenheim Foundation, the Fulbright Commission, the Social
Science Research Council and the National Science Foundation. In 2005
the American Sociological Association Methodology Section established
the Leo A. Goodman Award to recognize contributions to sociological
methodology, and/or innovative uses of sociological methodology, made by
a scholar who is no more than fifteen years past the Ph.D.
\end{abstract}


\end{frontmatter}

The~following conversation took place on January 10, 2008, at Leo
Goodman's home in Berkeley, California.

\textbf{Becker:} Leo, you frequently refer to me as your academic
grandson, as you were the thesis advisor of my thesis advisor, the late
Clifford Clogg. And, over the years, as good grandfathers do, you have
told me many stories about the people whom you have had the pleasure of
learning from and working with. Who got you started in thinking about
mathematics, about a career as a statistician and also about a career as
a statistician bridging sociology and statistics, and how did all this
come about?

%f1 ###
\begin{figure}[b]

\includegraphics{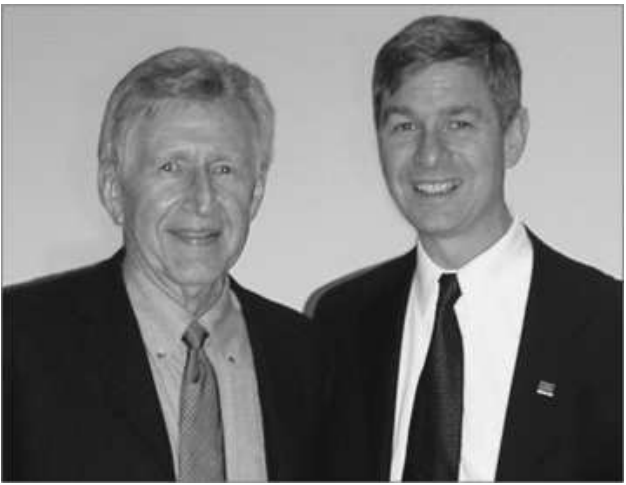}

\caption{Leo Goodman and Mark Becker.}
\end{figure}

\textbf{Goodman:} Mark, yes, I~do think of you with pride as a second
generation academic descendant of mine, and I would also like to say
here that I could think of myself as, in a certain sense, a first
generation academic descendant of my first real mathematics teacher,
Lipman (Lipa) Bers. He had been a student of Charles Loewner at the
Charles University in Prague, Czechoslovakia, and, during my
undergraduate years at Syracuse University, both Bers and Loewner were
faculty members in the Math Department there, having extricated
themselves from Europe just one step ahead of the Holocaust. I~took some
courses with Bers and with Loewner. (By the way, Bers once told me, many
years after I had been one of his students, that he and Loewner were
direct academic descendants of Gauss; Bers was a sixth generation and
Loewner a fifth generation descendant of Gauss. Also, it turns out that
Bers was a fourth generation and Loewner a third generation descendant
of Weierstrass.) Both Bers and Loewner were outstanding teachers, and
each of them produced mathematics of top quality.

I should stop here right now to say that I am partly joking when I say
that I could think of myself as, in a certain sense, a first generation
academic descendant of Bers. [Smile/Laughter] Bers was an important
person in my life, but he was not the advisor on my Ph.D. thesis; and,
as you know, being the Ph.D. thesis advisor is, strictly speaking, the
genealogical criterion that defines this kind of kinship. As you of
course also know, Sam Wilks and John Tukey were the ones who approved my
Ph.D. thesis.

%f2 ###
\begin{figure}[b]

\includegraphics{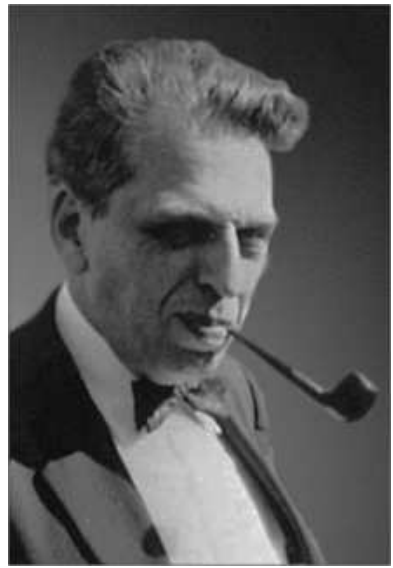}

\caption{Lipman Bers.}
\end{figure}

But now let me return for a moment to Bers and Loewner. Here is a brief
description of Loewner written by Bers, in words that could be used to
describe Bers as well:

``He $\ldots$  was a man whom everybody liked, perhaps because he was a man at
peace with himself. He conducted a life-long passionate love affair with
mathematics$\dots.$ His kindness and generosity in scientific matters,
to students and colleagues, were proverbial. He was also a good
storyteller, with a sense of humor$\dots.$ But first and foremost he
was a mathematician.''

%f3 ###
\begin{figure}

\includegraphics{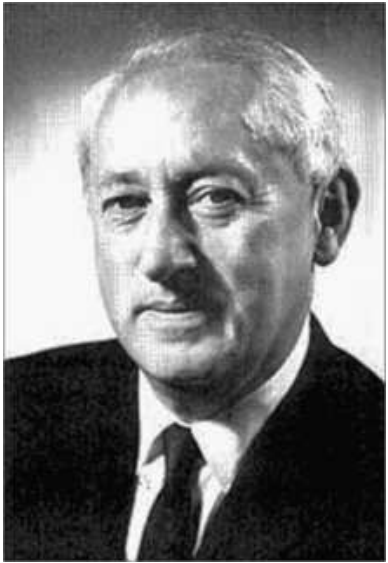}

\caption{Charles Loewner.}
\end{figure}

Mark, you asked me who got me started in thinking about math. I~would
say that Bers and Loewner were responsible for that. But, before them, I~would say that I got started in math because I actually got started as
an undergraduate major in sociology. [Smile/Laughter] During my
undergraduate days at Syracuse, sociology majors were required to take
the course in statistics given by the Sociology Department, and the
sociology faculty member who was assigned to teach the course was Robert
E. L. (Bob) Faris. He taught this course for the first time the year
that I took it. When he entered the classroom the first day of class, he
announced that he was assigned by the Sociology Department to teach this
course because he had written a book a long time ago that had some
tables in it, and also because he did know a little bit about
statistics, but he confessed that he didn't really know very much about
statistics, and he hoped that students in the class would be able to
help him to get through the course. Well, it happened to turn out that I
was able to help out. When the course came to the end, Faris told me
that he thought that I had a talent for statistics, and, if I would like
to gain still more strength in that subject, he would suggest that I
first strengthen my mathematics background by taking some courses in the
Mathematics Department. And so, that is what I then did.

Later on, when I was in my undergraduate senior year, it turned out that
I had taken just enough courses in math and just enough courses in soc
to graduate as a joint math/soc major. But what would I do when my
senior year would come to an end? I didn't know.

Bers then suggested to me that I should apply for graduate study in
mathematics at Princeton University, and Faris suggested that I should
apply for graduate study in sociology at the University of\break Chicago.
Bers, when he made his suggestion, also told me that no mathematics
undergraduate from Syracuse University had ever been accepted for
graduate study by the Mathematics Department at Princeton; and, with
respect to Faris' suggestion, I~was aware of the fact that he had been a
graduate student in sociology at the University of Chicago, and his
Ph.D. thesis, when it was published, turned out to be a kind of
sociological classic. So I applied for graduate study in math to
Princeton, and for graduate study in soc to the University of Chicago.

Now let me tell you a bit about Faris. He was a strong social
psychologist, and he was also very much a genuine sociologist. In both
spheres, his many contributions were informed by his strong commitment
to \mbox{sociology} as a discipline. In addition, he was an accomplished
painter, a pretty good violinist and an enjoyable pianist. We became
good friends.

Faris was a member of a four-generation line of sociologists. His
father, Ellsworth Faris, had served for fourteen years as the first
chairman, after the founding chairman, of the University of Chicago
Sociology Department, and he ranks high in the final hierarchy of those
who achieved major results in the building of American sociology during
the first half of the twentieth century. The~father served as President
of the American Sociological Association in 1937, and the son in 1961;
the father served as Editor of the \textit{American Journal of
Sociology}, and the son as Editor of the \textit{American Sociological
Review}. Bob Faris' son, Jack, received his Ph.D. in sociology from the
University of Chicago; and Jack's son, Robert W., received his Ph.D. in
sociology from the University of North Carolina at Chapel Hill. Jack is
now the President of the Washington Biotechnology \& Biomedical
Association (earlier he had been the University of Washington Vice
President for University Relations); and Robert W. has just now
completed his first year as an Assistant Professor in the Sociology
Department at University of California at Davis.

\textbf{Becker:} Well, Leo, Faris gave you rock solid advice for
launching a career in sociology, but you ultimately followed Lipman
Bers' suggestion and went on to Princeton to study mathematics. How did
you decide to do that, and how did you decide to emphasize mathematical
statistics while you were a graduate student in mathematics?

\textbf{Goodman:} Well, after I had mailed out the application to
Princeton in math and the application to the University of Chicago in
soc, I~did wonder what would happen next.

Sometime during my senior year, when I was visiting my parents in New
York, I~decided to take the train from New York to Princeton Junction, just in
order to see what the Princeton campus looked like. After the train
arrived in Princeton Junction, I~walked from there to the campus, and
then walked around the campus, and was impressed by how beautiful it
was. Then I just happened to come across Fine Hall, the Mathematics
Department building, and it too was beautiful. It was located in the
southeastern corner section of the campus, and it harmonized with the
other structures in this, the ``red brick section'' of the campus. Red
brick and limestone were used in the Collegiate Gothic architectural
style of these structures, and they presented a unified appearance. I~was also aware of the fact that, when the Institute for Advanced Study
was first founded in Princeton, it had temporary quarters in Fine Hall.
All this left its impression on me.

Fine Hall was in the shape of a box (a rectangular parallelepiped).
[Smile] The~hallway inside the building was rectangular in shape, and it
went around the inside of the building, with rooms on each side of the
hallway. I~walked around the hallway, and then I walked around it a
second time, and maybe a third time. (Einstein, I~thought, must have
worked at an earlier time in one of the offices located right off this
hallway.) Then, a secretary, whose office door had been open, noticed
that there was this strange young man (me) walking around the hallway,
and she came out of her office and asked me if she could be of any help.
I~told her that I was an undergraduate senior at Syracuse University,
and had applied to Princeton for graduate study in math. She then asked
me to wait there for a moment, and she went into another office, which
was adjacent to her office. Then she came out of the other office with a
man who said he was Sam Wilks.

%f4 ###
\begin{figure}

\includegraphics{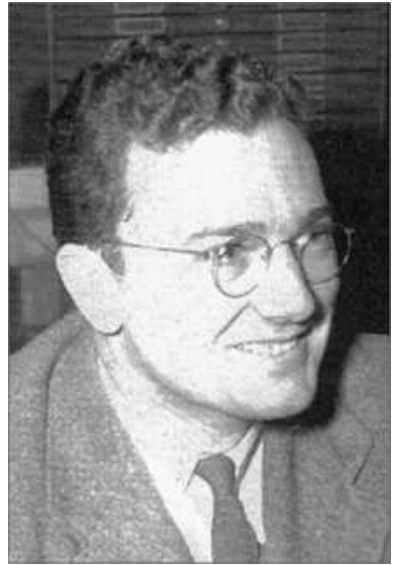}

\caption{Sam Wilks.}
\end{figure}

He invited me into his office, which was also beautiful. It was
spacious, with carved oak paneling and a splendid fireplace. The~office
contained a large conference table with chairs all around it, and also a
large desk, and wooden bookshelves filled with books from floor to
ceiling. Wilks asked me to sit down, and we then talked for maybe an
hour or more. He had a very pleasant Texan drawl. After about an hour or
more, I~began to think that I must be taking up too much of his time,
and I got up to leave; and he asked me to wait a minute, he wanted to
first make a phone call. He then phoned a fellow faculty member, Fred
Stephan, who was a distinguished sociologist, statistician and
demographer in the Sociology Department there, and he told Stephan over
the phone that an undergraduate senior at Syracuse University who had
applied to become a graduate student in the Mathematics Department at
Princeton was now in his office, and he thought that Stephan would be
interested to meet this student. He then gave me directions on how to
get to Stephan's office, and off I went. I~found Stephan's office, and
he and I also had a very nice conversation.

After all this, I~felt elated, and I thought that, if it turns out that
I am accepted by the Math Department here at Princeton, here is where I
will go. I~then started walking back to Princeton Junction, walking on
cloud nine, and I got on the next train in order to return to New York.
But I soon realized, after the train had pulled out of the train-station
and was on its way, that it was heading in the wrong direction, heading
for Philadelphia. [Smile/Laughter]

So that's how I made the decision to be a graduate student in math if it
turns out that I am accepted at Princeton, and not a graduate student in
soc at the University of Chicago.

By the way, Mark, I~have been telling you here about my feelings and
thoughts while on the trip that I made from New York to Princeton just
in order to see what the campus looked like. Please keep in mind that I
was nineteen years old at that time (and it is now about sixty years
after the events that I have been describing), and I am trying to convey
to you, as honestly as I can, what that nineteen-year-old felt and
thought at that time.

Yes, I~made the trip to Princeton just in order to see what the
Princeton campus looked like, and it just happened that I came across
Fine Hall, and when I was walking around and around in awe in the Fine
Hall hallway, the secretary whose office door just happened to be open
just happened to come out of her office and just happened to ask me if
she could be of any help, and she just happened to be Wilks' secretary.
Mark, it seems to me that all these happenings might be viewed as
examples of what one might call dumb luck. I~have always thought of
myself as a very lucky person, and I feel deeply grateful for all the
lucky events that have occurred to me in my life. Just imagine what
might have happened if, when I was walking around and around in the Fine
Hall hallway, the secretary whose office door was open was the secretary
of, say, Solomon Lefschetz, who was the math department chairman at that
time. (I~will say more about Lefschetz later.) If that had happened, I~suppose it is possible that I might have ended up deciding to be a
graduate student in sociology at the University of Chicago. [Smile]

\textbf{Becker:} Leo, let's talk about your student experiences at
Princeton. Wilks and Tukey were your major professors at Princeton. What
are your remembrances of these two luminaries of statistics, and their
respective contributions to your development as a scholar?

\textbf{Goodman:} Let me begin by telling you about two very nice
experiences that I had with Tukey during my first year as a graduate
student: As I said earlier, when I graduated from Syracuse, it turned
out that I had taken just the smallest number of courses in math and the
smallest number of courses in soc to graduate with a joint major, and I
had the impression, after being a graduate student at Princeton for just
a short time, that all of the other first year math graduate students
had been studying mathematics intensively full time when they were
undergraduates, and also possibly when they were in high school, and
maybe even when they were in elementary school. [My cohort of graduate
students in the math department included the future Nobel Laureate John
Nash (the ``Beautiful Mind'') and many other brilliant students.] One
day, sometime after I had been at Princeton for maybe two or three
months, I~happened to be walking in the Fine Hall hallway, and Tukey
happened to be walking in the hallway too, and our paths happened to
cross. He stopped me and asked, ``How are you doing?'' I then said, ``I
don't know how I'm doing.'' He then said, ``Follow me.'' He opened the
door of an empty classroom, and he asked me to go up to the blackboard
at the front of the room, and he took a seat near the back of the room.
He then asked me a math question to explain something or other, and I
tried to answer the question writing on the blackboard. I~could see
that, while I was trying to answer his question at the blackboard, he
was doing something else seated there near the back of the room,
possibly writing an article. (It was well known that Tukey was able to
do two things at the same time.) When I finished trying to answer the
first question, he didn't comment on my attempted answer, and he asked
me a second math question. I~then tried to answer that question writing
on the blackboard. Then a third math question; and the questioning and
attempted answering continued for maybe an hour or more. Finally the
questioning stopped. Tukey then got up from his seat near the back of
the room, he didn't say anything, and he walked very slowly toward the
front of the room. The~expression on his face was that of a man thinking
about serious matters. He had his hand on his chin, which was a typical
place where he put his hand when he was thinking about serious matters.
And this is what he finally said to me, speaking very slowly: ``Well,
[long pause] what I think you really need, [extra long pause] is some
folk dancing.''

%f5 ###
\begin{figure}[b]

\includegraphics{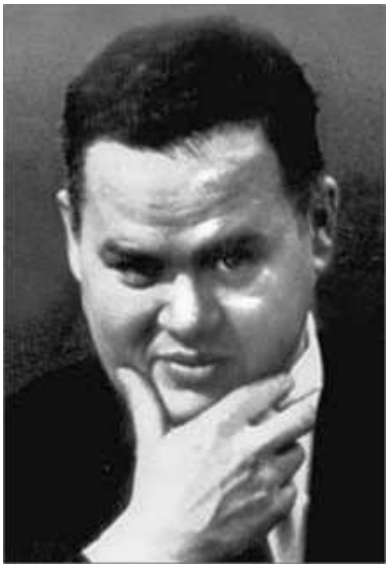}

\caption{John Tukey.}
\end{figure}

Mark, this comment by Tukey is a good example of his special kind of
sense of humor and his at times elliptical manner of speech. He was
telling me, in his own way, that I was doing fine, that my ability to
answer math questions was fine, and that I ought to take time out for
folk dancing or for whatever else might please me. He then told me when
and where the folk dancing would take place, and he invited me to come
to it, which I then did. (Mark, by the way, when Tukey invited me to
come to the folk dancing, he didn't tell me that the folk-dancing group
met under his direction, that he was the folk-dance leader, and that he
and some of his friends demonstrated folk-dancing steps for beginners,
which is what I was.) Later that year, Tukey met his future wife,
Elizabeth Rapp, at a folk dancing session.

Here now is the second very nice experience that I had with Tukey during
my first year: Attached to the wall in the hallway just outside the Math
Department office in Fine Hall were the mailboxes for the math faculty
members. One day, sometime after I had been at Princeton for maybe four
or five months, I~happened to be walking past the math office, and Tukey
happened to be standing there reading a postcard that he had just taken
out of his mailbox. Written on the postcard was a statistical problem
sent to Tukey by someone named Allen Wallis. (I will say more about
Wallis later.) Tukey handed me the postcard, and then said something
that I didn't understand about the statistical problem written on the
postcard. He said just one sentence about the problem. Without him
explicitly saying so, I~realized that he wanted me to work on the
problem that was described on the postcard. (Tukey was a New Englander,
and he spoke with a sort of ``down-east'' accent. He also spoke in an
elliptical, enigmatical, oracular fashion, and he liked to coin his own
words.)

I then went off with the postcard in order to begin to think about the
statistical problem and about what Tukey had meant in his one sentence
about the problem. After thinking about this for a while, I~found that I
still just couldn't figure out what he had meant, and I finally just
gave up trying to do so. But I didn't give up working on the problem. I~worked on this for maybe three or four weeks, and then I wrote up the
results that I had obtained in the form of a paper, but without putting
an author's name on it. And I put the paper in Tukey's mailbox. A day or
so later, Tukey asked me to come by his office. When I came to his
office, he told me that this work was very good, and that it should be
published as is in the \textit{Annals of Mathematical Statistics} right
away. He started to write my name on the paper as the author, but I said
no, that he and I were the authors---he gave me the problem and he gave
me his help. He said no, that I was the author, and this was my work.

All this happened sometime near the end of 1948 or the beginning of
1949, and my paper was published in the \textit{Annals of Mathematical
Statistics} in December 1949.

Now, let me tell you about Wilks. As I said earlier, it was after I
first met Wilks, and had a chance to talk with him in his office, that I
thought that, if it turns out that I am accepted as a graduate student
by the Math Department at Princeton, that is where I will go. I~will
give up the notion of becoming a graduate student in soc at the
University of Chicago.

Mark, as you of course know, Wilks was the father of mathematical
statistics at
Princeton and a major leader in the development of this discipline; and
many of the people who had taken their Ph.D.'s under Wilks (for example,
Fred Mosteller, Ted Anderson, Don Fraser, Ted Harris, Will Dixon, Alex
Mood) also had important roles in this development. Wilks was very
friendly and very fair. Everyone liked him. He was a quiet, penetrating
and influential leader in the work of many organizations, especially in
mathematics, statistics and social science. To these organizations, he
brought wisdom, commitment and persistence. He had a remarkable sense of
what was important and what was not. One of the many important
contributions he made to statistics was his work, for a period of about
thirteen years, as the first editor of the \textit{Annals of}
\textit{Mathematical Statistics}, when it became a publication of the
Institute of Mathematical Statistics (IMS). (He was also one of a small
group of statisticians who founded and organized the IMS, and, from its
inception, he was a leading member of this organization.) During the
period of Wilks' editorship of the \textit{Annals}, he turned it into
the foremost, internationally recognized, journal of mathematical
statistics; and this had an important influence on the subsequent
development of the field of statistics.

I mentioned here Wilks' editorship of the \textit{Annals} in part
because, during my first year and a half at Princeton, Wilks would from
time to time give me a research manuscript, which had been submitted to
the \textit{Annals}, for me to referee; and I think that this experience
of refereeing manuscripts really contributed to my education and
training as a statistician.

Sometime during the first half of my second year at Princeton, Wilks
asked me to select any, more or less, contemporary statistics article,
by any statistician, and then study the article and give a talk on it,
explaining the results that were in the article. I~don't recall how I
went about trying to select an author and an article, but it turned out
that I selected a recently published article by Charles Stein. (Stein
had received his Ph.D. at Columbia under Abraham Wald about two years
earlier, and his articles, which he had written during the period
starting from two years before he received his Ph.D. until the time when
I was trying to select an article, were all very exceptional.) I studied
the selected article and prepared the talk that I would present, but
there was a small part of one section in the article that I just
couldn't understand, and I decided to leave that section out of my talk.
On the day that my talk was scheduled to take place, I~entered the
lecture room at the appropriate time, the members of the audience were
in their seats, and just as I was about to begin my presentation, in
walked Wilks and Tukey escorting a third person I didn't know. The~three
of them sat down in the last row of the lecture room. (The~third person
turned out to be Willy Feller, a distinguished mathematician
specializing in probability theory, who was at that time a professor at
Cornell University, and was visiting the Princeton Math Department for a
few days. Wilks, Tukey and the Math Department were trying to persuade
Feller to leave Cornell and become a professor in the Princeton Math
Department.) I presented my talk, and, immediately after it was over,
Feller came rushing right up to me. He introduced himself, and, with a
really big smile on his face, he told me that he really enjoyed the talk
and found it to be very interesting. Well, Mark, I~was of course pleased
that Feller liked the talk, and I also was pleased when I found out a
week or so later that, after his short visit to Princeton, Feller did
decide to join the Princeton Math Department.

By the way, many years later, in a conversation that I had with Charles
Stein, I~told Charles that I just couldn't understand a small part of
one section in the article of his that I had studied when I was a
graduate student, and he then told me that there was a mistake that he
had made in that section.

Now, back to Wilks. What I am about to tell you happened, I~think, a few
years after I had received my Ph.D. degree, but it is possible that it
happened sometime before I received the degree: I received a letter from
the University of Texas informing me that Wilks was being considered for
an appointment as the President of the University, and asking me to send
them a letter of reference about Wilks. Well, I~wrote the most
affirmative thumbs-up letter that I have ever written about anyone. And
I remember that it gave me great pleasure to be able to do this. It then
turned out that Wilks was invited to become the President of the
University, he did consider the offer, but in the end he turned it down.

Mark, earlier I had mentioned to you that Wilks had a pleasant Texan
drawl. But there is more to say about Wilks, who was very much a Texan,
turning down the Presidency of the University of Texas: Wilks was born
in a small town bordering on a nice lake in North East Texas, and he was
raised with his two younger brothers on a 250 acre ranch, which his
father farmed, outside of the small town. Wilks and his brothers and his
parents had very close family ties, and, after he was settled in
Princeton, he would take advantage of whatever opportunities would turn
up for him to revisit Texas and visit with his family. Also, Wilks' son,
Stanley, would go to Texas whenever possible (for example, during his
summer vacations from school) in order to visit with his two uncles and
their families, and with his grandparents; and, when Stanley was a
senior in high school in Princeton and was considering where to go to
college, he decided to go to Texas as an undergraduate at North Texas
State college, the same undergraduate college that his father had
attended, and not too far away from where his two uncles and their
families, and his grandparents, lived. So you can imagine how surprised
and disappointed Wilks' family must have been when he told them that he
had decided to turn down the job offer and remain in Princeton.

Here now is a story about how Wilks' family viewed all this. I~can't
remember who told me this story, and I can't vouch for its veracity. I~am telling this to you because I think it is amusing: Wilks' two
brothers had good jobs in Texas, and his parents thought that, with
three grown-up sons, it isn't so unusual that one of the sons might just
not be as able as the other two, and just might not be able enough to
obtain a good job in Texas. When the parents first heard that Wilks had
been offered this job at the University of Texas, they were very pleased
to hear the news, and they could then see that this son too was able
enough to obtain a good job in Texas. We are then left to speculate
about what Wilks' parents thought about his decision to turn down this
offer of a good job in Texas.

\textbf{Becker:} Princeton has long had one of the world's finest
mathematics departments. You were fortunate to have studied there with
statistical greats like John Tukey and Sam Wilks, but what about your
interactions with mathematics professors who were not statistically
inclined? Do you have any particular remembrances of those?

\textbf{Goodman:} Math graduate students at Princeton at the time when I
was there were not required to attend any courses. All you had to do was
pass an oral exam, called the general exam, covering four subfields of
math, usually taken when your first year as a graduate student was
completed, or sometime after that. You then had to submit a thesis, and
have the thesis approved. Also, there was a foreign language
requirement, two foreign languages of your choice, and, for each of
these languages, you had to demonstrate to a math faculty member of your
choice that you had a reasonable ability to read ordinary mathematical
texts that were written in the foreign language. (There seemed to be a
general understanding among the math graduate students that the math
faculty didn't take the language requirement very seriously.) As a math
graduate student at Princeton, you had the feeling of having almost
complete freedom.

Although it wasn't required, I~actually did take some courses during my
first year as a graduate student: I~remember taking a course in analysis
given by Salomon Bochner, and a part of a course in mathematical logic
given by Alonzo Church, and I think that I also may have taken a course
in point-set topology given by Ralph Fox, but I am not sure about that. I~remember that I was very impressed with both Bochner and Church, by
their great skills as teachers, by the depth of their understanding of
their subjects, and also by their unforgettable personalities. (If I did
attend Fox's course, I~just don't remember what it was like. I~will tell
you a little bit about Fox a little later.)

Bochner was one of the foremost twentieth-century mathematicians whose
research profoundly influenced the development of many different areas
of analysis. He was born into a Jewish family in what was then a part of
Austria--Hungary, and is now a part of Poland. Fearful of a Russian
invasion at the beginning of World War I, his family moved to Germany,
seeking greater security. Bochner was educated at the University of
Berlin, where he wrote his dissertation, and he then lectured and made
very important contributions on a surprising variety of topics in
analysis in Germany, at the University of Munich, for about ten years.
His academic career in Germany came to an abrupt end a few months after
the Nazis came to power, when laws were established providing for the
removal of Jewish teachers (and those of Jewish descent---those having
at least one grandparent of Jewish descent) from the universities, and
he then received a timely offer of a position at Princeton, which he
accepted. He was a very important faculty member at Princeton for the
next 36 years.

%f6 ###
\begin{figure}

\includegraphics{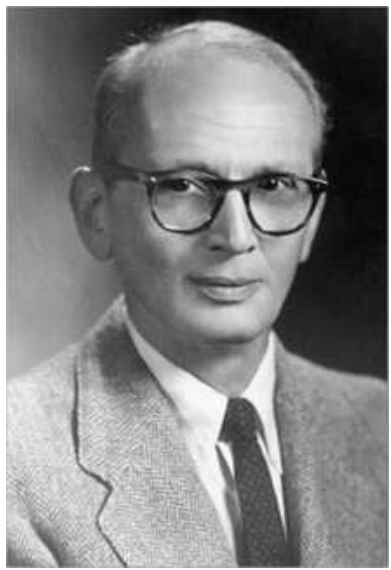}

\caption{Salomon Bochner.}
\end{figure}

Mark, now you might find it interesting to compare what I have just now
told you about Bochner with what I will now tell you about Alonzo Church
and his pedigree: Church was a mathematician who was an early pioneer in
and major contributor to the field of mathematical logic, and he was
also responsible for some of the foundations of theoretical computer
science. His great-grandfather, Alonzo S. Church, was a professor of
mathematics and astronomy, and then became the president of the
University of Georgia for a period of thirty years. Alonzo Church's
grandfather, Alonzo W. Church, was at one time Librarian of the U.S.
Senate. His father, Samuel R. Church, was a Justice of the Municipal
Court of the District of Columbia, but he resigned from that post
because of failing vision and hearing. The~family then moved to rural
Virginia, where Alonzo Church and his younger brother grew up. He was an
undergraduate at Princeton, graduating with an A.B. in mathematics, and
then he continued as a graduate student, completing his Ph.D. there.
Following a two-year post-doctoral National Research Fellowship, which
he spent at Harvard University the first year, and the University of
Gottingen and the University of Amsterdam the second year, he was
invited to return to the Princeton Math Department, to begin his
academic career there. He was a very important faculty member at
Princeton for the next 39 years.

%f7 ###
\begin{figure}

\includegraphics{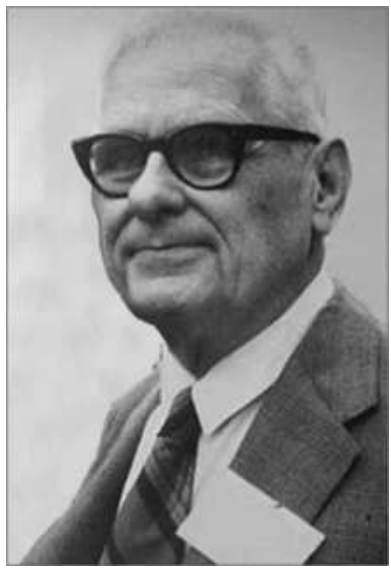}

\caption{Alonzo Church.}
\end{figure}

Here now is a nice story that I would like to tell you about Bochner:
Sometime during my first year as a graduate student, during the period
of time when I was taking his course, he once told me that, if he had
his life to live all over again, he definitely would not choose to be a
professor. So I asked him what would he choose to be instead. He then
told me that he would choose to be a laundry-truck driver. I~then asked
him why would he make that choice. ``Well,'' he said, ``when you're a
laundry-truck driver, you drive the truck to the first house on the
delivery schedule, then you deliver packages of clean diapers to the
housewife, and you pick up packages of dirty diapers from her, then you
drive the truck to the second house on the delivery schedule, and you
repeat the same procedure there, then you drive to the third house $\ldots$
and you continue to do this all day long, every working day.
\textit{And}, while you are doing this during all that time,
[extra long pause] \textit{you can also simultaneously spend all
that time proving interesting theorems!!!''} [Smile/Laughter] Bochner
was a civilized and erudite mathematician, also a lover of music and
art.

Mark, earlier in our conversation, I~told you that there was a language
requirement, two foreign languages of your choice, and, for each of
these languages, you could ask a math faculty member of your choice to
test you on that language. I~selected French and German, and Bochner
passed me in French, and Solomon Lefschetz passed me in German. I~have
just now been telling you here about Bochner, and I would like now to
tell you about Lefschetz. (Mark, before we move on, I~should say here
right now that, with respect to the language requirement, if the
pass/fail decision had been in the hands of the corresponding Princeton
language departments, instead of the math department faculty, I~would
have had a much harder time trying to pass the language requirement.)

Now about Lefschetz: When he was 23 years old, he was working as an
electrical engineer in the U.S., and, in a terrible industrial accident---a transformer explosion---he lost both his hands and a part of each
forearm. For the rest of his life, he used a pair of artificial hands
(wooden hands, gloved)---his prosthetic hands---that fit over the
remaining parts of the forearms. Three years after the accident, he
enrolled as a doctoral student in mathematics at Clark University in
Massachusetts, and he received his Ph.D. a year later with a thesis on
algebraic geometry. In the next thirteen years, he produced many
research articles of striking originality and importance in algebraic
geometry and algebraic topology. He then was invited to be a visiting
professor at Princeton, and, at the end of his first year there, he was
offered a permanent post, which he accepted.

At Princeton, he profoundly affected the development of mathematics in
the U.S. as the editor of the \textit{Annals of Mathematics} during a
thirty-year period. During his editorship, the journal became one of the
principal journals of research mathematics in the world. Also, during
Lefschetz's chairmanship of the department, eight new faculty members
were appointed, all of first rank. A mathematics faculty of first rank
became an even larger faculty of first rank, one of the world's great
centers of mathematical research and teaching.

%f8 ###
\begin{figure}

\includegraphics{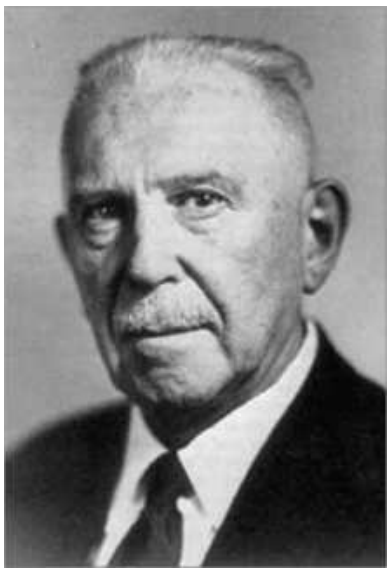}

\caption{Solomon Lefschetz.}
\end{figure}

Lefschetz was born in Moscow into a Jewish family (his parents were
Turkish citizens), and soon after his birth they moved to Paris. He was
educated there in engineering, and he emigrated to the U.S. when he was
21. Two years later, there was that terrible industrial accident. When
he first came to Princeton in the 1920's, he was one of the first Jewish
faculty members on the Princeton campus, and it is reported that he felt
that people avoided him in the hallways and on campus on that account.
He said that he was an ``invisible man'' there at the time. It is also
the case that he could be rude, imperious, idiosyncratic and
obstreperous, with a commanding (bossy) personality. He also was a man
who had a really great amount of energy, a supercharged human
locomotive.

Lefschetz was sometimes accused of caving in to anti-Semitism at
Princeton for refusing to admit many Jewish math students---his
rationale being that nobody would hire them when they completed their
degrees. (It is my impression that Lipman Bers was implicitly alluding,
in part, to this kind of problem at Princeton when he told me that no
mathematics undergraduate students at Syracuse University had ever been
accepted for graduate study by the Mathematics Department at Princeton
and he suggested to me that I should apply for graduate study there.)
Times have really changed at Princeton since the bad old days.

\textbf{Becker:} Leo, many academics have rich or humorous stories
relating to the exams they experienced en route to the Ph.D. When did
you take your oral general examination, and what was that like?

\textbf{Goodman:} I took my oral general exam soon after the very
beginning of my second year as a graduate student. This exam covers two
required subjects, algebra and real and complex variables, and two
special advanced topics of the examinees choice; mathematical statistics
and point-set topology were what I chose.

The~examinee is not told beforehand who his examiners will be. He finds
out when he enters the examination room and sees the examiners sitting
there. When I opened the door to the room, sitting there were Salomon
Bochner, Emil Artin, Ralph Fox and Sam Wilks.

I haven't yet told you anything about Artin and Fox. First, about Artin:

He was one of the leading algebraists of the twentieth century. He was
brought up in a town that was mainly German speaking, in what was then a
part of the Austrian Empire, and is now in the northern part of the
Czech Republic. He received his doctorate in mathematics in Germany,
from the University of Leipzig, and he began his academic career at the
University of Hamburg. He lectured and made many very important
contributions to a wide range of topics in algebra there over a period
of eleven years before Hitler came to power. His position at the
university was not affected by the laws established a few months after
Hitler came to power providing for the removal of Jewish teachers from
the universities, since he wasn't Jewish. However, since Artin's wife
was Jewish, his position at the university was affected four years later
by a new law that provided for the removal from the universities of
those teachers related to Jews by marriage. He then came to the United
States, taught at Notre Dame for one year, at Indiana University for
eight years, and then at Princeton.

%f9 ###
\begin{figure}

\includegraphics{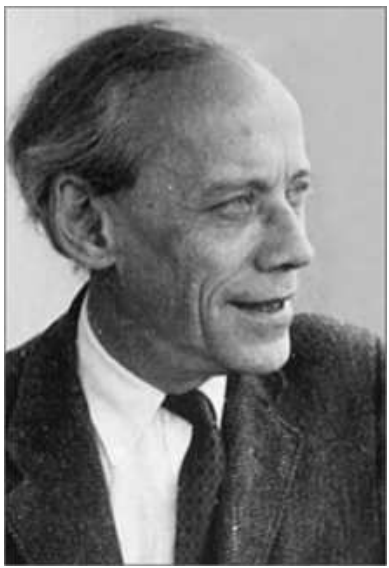}

\caption{Emil Artin.}
\end{figure}

Now about Fox: He was an American mathematician who devoted most of his
career to the field of topology, and, in particular, to knot theory. He
taught and advised many of the later contributors to topology, and he
played an important role in the modernization and development of knot
theory. After receiving his Ph.D. from Princeton, he spent the following
year at the Institute for Advanced Study in Princeton, and he then
taught at the University of Illinois and Syracuse University before
returning to Princeton, six years after having received his Ph.D. there,
to join the math faculty.

One of Fox's strong interests was the ancient Japa\-nese board game of Go.
He represented the United States in the first international Go
tournament, held in Tokyo, and he popularized the playing of Go at the
Princeton Math Department. Go was one of two favorite board games that
were played in the Fine Hall common room just about every day by some of
the math graduate students at tea time and at other times too.

Sometime before the time when I took the general exam, I~was sitting
around in the Fine Hall common room, chatting with another graduate
student who was then in his third year at Princeton and who had taken
his general exam the year before, and he asked me how was I preparing
for the part of the exam on complex variables. So I told him what I had
been studying in complex variables, and he asked me what did I know
about complex manifolds. I~said that I didn't know anything about
complex manifolds. He then offered to tell me about it. So the two of us
went into one of the empty classrooms, I~sat down, and he went to the
blackboard and started to tell me about complex manifolds writing on the
blackboard. After about an hour or so of his exposition, I~thought that
that was enough, and we stopped.

Well, sometime later, on the day when I went into that exam room, and
was facing the four examiners, the first one to ask me a question was
Bochner. He said: ``What do you know about complex manifolds?'' I then
said, ``Not very much.'' And he said, ``Tell me what you do know about
complex manifolds.'' So I went up to the blackboard and proceeded to
write on the blackboard (facing the blackboard), trying to repeat what I
had been told earlier about complex manifolds by the more advanced
graduate student. After a while, as I was proceeding with the
exposition, I~suddenly heard someone (one of the examiners), with an
authoritative-sounding voice, say, ``\textit{That is
incorrect!}'' I then turned from facing the blackboard to facing the
examiners, and I could tell that the examiner with the authoritative
voice was Artin. Then Bochner said, ``No, that is correct.'' And the two
of them, Artin and Bochner, then started to argue with each other as to
whether what I had said was incorrect or correct. Meanwhile, I~backed
up, and leaned against the blackboard until the argument came to an end.
After that, Artin, Fox and Wilks, each of them in turn, asked me his own
questions, and I did my best to try to answer them correctly.

Then I was asked to leave the exam room and wait outside the room. After
a little while, Wilks came out with a nice smile on his face, and he
congratulated me, and said that in his experience he'd never seen an
examinee answer questions so calmly, and that I did a nice job. Well,
Mark, I~was of course very pleased to hear this, especially since, when
I left the exam room, I~wasn't sure that I had passed the exam! [Smile]

Thinking some more about Wilks' comment about examinees answering
questions calmly, I~am reminded of something amusing that Wilks' wife,
Gena, had told me about a former Wilks student and his general exam.
This conversation with Mrs. Wilks took place sometime near the end of my
first year as a graduate student at a time when I was supposed to be
studying in preparation for my general exam. My studying was interrupted
by a sudden appendicitis, with surgical removal of the inflamed appendix
carried out just in the nick of time, and with a required stay in the
hospital for a few days to recover from it all. Mrs. Wilks came by to
visit with me there, and she asked me how I was feeling. I~told her that
I was recovering just fine, but I wasn't able to focus on my studies for
the general exam. She then told me this story: There was this
well-known, very good statistician, a former very good Wilks student
(whom I will call student X), who, a few days before he was scheduled to
take the general exam, came into Wilks' office, and said the following
to Wilks, ``Sam, I~don't know a thing! I'm not going to take the exam.''
So Wilks told student X that the exam would be put off for two months.
Then, two months later, a few days before student X was supposed to take
the rescheduled exam, he again came into Wilks' office, and this time he
said the following: ``Sam I still don't know a thing! I'm not going to
take the exam.'' So Wilks told student X that the exam would be put off
this time for one month. Then, one month later, a few days before
student X was supposed to take this rescheduled exam, he again came into
Wilks' office and again said the same thing that he had said twice
before. Wilks then told student X that he should take it easy, go out to
a movie or do something else that would be relaxing and fun, and that he
should drop by Wilks' office the next day at 10 AM. Student X then came
by Wilks' office the next day at the appointed time, and he found the
four-member examination committee ready to proceed with the exam.
Student X passed the exam, and then went on to become a well-known, very
good statistician.

\textbf{Becker:} Once you had your examinations behind you, how did you
obtain a dissertation topic, and how did you proceed to write your Ph.D.
thesis?

\textbf{Goodman:} After my general exam was over, I~star\-ted to try to
think about possible thesis topics. One day, while I was thinking about
statistical problems that had been discussed in manuscripts that I had
refereed earlier for Wilks, an idea just happened to come to me by free
association, and with this idea there came a statistical problem that I
thought I would try to work on. I~then worked on this problem for a few
weeks, but I didn't seem to be getting anywhere with the problem. Then,
it crossed my mind that, during the time period when I slept each night, I~may have been dreaming about the problem that I had been working on
during the daytime; but, if I had been dreaming, when I awoke each
morning the dream was gone. So I put a pad of paper and a pencil on the
night-table next to my bed, and that night, I~awoke in the middle of the
night and proceeded to write down on the pad of paper what I had been
dreaming---a very long dream. I~then fell back to sleep, and I slept
for a long time. When I awoke, I~noticed the pad of paper lying next to
me on the bed with a lot of writing on it, and I wondered what the
writing was about.

After looking over the dream document, I~could make out that it was
about the problem that I had been working on during the daytime. I~could
make out, in some of the paragraphs, what were the ideas or results
contained in them; and in some of the paragraphs, I~couldn't. Well, I~then worked on the dream document for a few months, developing and
extending it; and, when I was finished doing that, I~wrote it all up in
the form of a Ph.D. thesis, and I put a copy in Wilks' math department
mailbox, and a copy in Tukey's. After a week or so, Wilks told me that
it was nice work and that he approved it, and Tukey told me that, first, I~should include in the thesis a numerical example that illustrates the
statistical method introduced and developed in the thesis; and, second, I~should give a talk on the thesis in August at the annual Summer
Seminar in Statistics, which would take place in Connecticut (it turned
out that Tukey was one of the organizers and leaders of the Summer
Seminar); and, third, he approved the thesis.

Sometimes, Mark, when someone happens to ask me about the writing of my
Ph.D. thesis, I~like to say simply that the thesis was given to me in a
dream. [Smile] It turns out that there is some truth in my saying that;
and, to the person asking me about the writing of the thesis, I~am
willing to clarify what I mean, if clarification is what is called for.
[Smile]

\textbf{Becker:} You opted to take your first faculty position at the
University of Chicago, and yet at that time there was not a formal
statistics department at Chicago. What or who motivated you to accept
the offer at Chicago?

\textbf{Goodman:} During the Winter/Spring of 1950, while I was
completing my work on the Ph.D. thesis, I~began to think about what I
might do when the thesis was done. A short time before that in 1949,
Allen Wallis (about whom I will say more later), who was at the
University of Chicago at that time, had persuaded the Chancellor of the
university to permit him to establish a ``Committee on Statistics,''
which was to be essentially a nascent department, but the Chancellor at
that time was unwilling to approve a ``Department of Statistics.'' When
I visited the University of Chicago to look it over, I~was intrigued by
the idea of being a part of a small group of statisticians that could
grow into a pride of statisticians, and I was also aware of the fact
that the Sociology Department at the university had been for many years,
and still was, a very distinguished department. I~also had the
impression that the University of Chicago was a good place to be an
assistant professor, and I liked the spirit of the place. So that is why
I opted to go there, rather than to one of the other good universities
where I might have gone.

\textbf{Becker:} Your research collaborations at the University of
Chicago with Bill Kruskal on measures of association for
cross-classifications spawned a citation classic. How is it that you and
Kruskal came to work on measures of association for a
cross-classifica\-tion of counts?

\textbf{Goodman:} Bill Kruskal and I arrived at the University of
Chicago at about the same time, in time for the beginning of the
1950--1951 academic year. We became colleagues and good friends, and
we\break
worked together very harmoniously and productively as colleagues, and
also as co-authors, over a very long period of time, even after we
completed our series of four joint articles and after Bill became the
Dean of the Social Science Division at the university, and even after I
left the university in 1987 to begin working at the University of
California, Berkeley.

%f10 ###
\begin{figure}[b]

\includegraphics{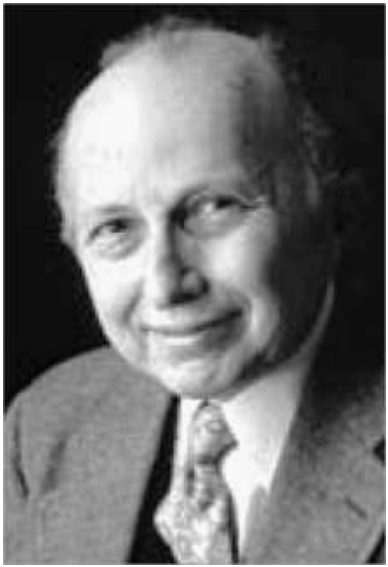}

\caption{Bill Kruskal.}
\end{figure}

Bill and I started to work together in the early 1950s on the
introduction and development of various measures of association for the
analysis of cross-classified categorical data, and we published our
first joint article on this subject in 1954, followed by a series of
three other joint articles on the subject in 1959, 1963 and 1972; and
the four articles were brought together in a single volume in 1979. Bill
and I worked on the first article---the core article---on and off for
about two years before we submitted it for publication, and the series
of four articles evolved over a 20-year period. The~1979 volume appeared
in print 25 years after the publication of the first article.

The~joint work that Bill and I did grew out of a conversation that we
had at a New Year's Eve party that each of us happened to attend at The~Quadrangle (Faculty) Club at the university. Our conversation at the
party was about our earlier experiences serving as statistical
consultants after we arrived at the university. As beginning faculty
members, Bill had been asked to serve as a statistical consultant to
Bernard Berelson in the Graduate Library School, and I had been asked to
serve as a statistical consultant to Louis Thurstone in the Psychology
Department.

Berelson was the Dean of the Graduate Library School at that time and
later became the President of the Population Council. He also was an
important figure in the social and behavioral sciences at that time, and
later he became an even more important figure. Thurstone was a
distinguished professor in the Psychology Department where he was the
founder and \mbox{director} of the Psychometric Laboratory. He had been
instrumental in the development of the field of psychometrics, and was
at that time the major figure in the development of factor analysis.

Well, the conversation that Bill and I had at that party took place
after Bill had met with Berelson and after I had met with Thurstone and
some other members of his Psychometric Laboratory. Bill and I were
describing to each other what happened when he met with Berelson and I
met with Thurstone and his group. And we observed in this conversation
that the kinds of statistical problems that Berelson was concerned with
and the kinds of problems that Thurstone and his group were concerned
with could be viewed as problems concerning the measurement of
association for cross classifications. We discovered that each of us had
been independently thinking about similar kinds of questions. So, right
then and there, at that party, Bill and I joined forces, and we were off
and running.

As I said earlier, Bill's and my first joint article---the core
article---was published in 1954. In 1979, the Institute for Scientific
Information (ISI) informed us that this article had been selected as a
Citation Classic, and we were invited to write a commentary on that
article, which the ISI published in \textit{Current} \textit{Contents,
Social and Behavioral Sciences.} It turns out that, according to the
ISI, our 1954 joint article has been cited about 1125 times so far.
This article still continues each year to be cited in a wide range of
different articles in journals that cover a very wide range of different
fields of study. In each of the past three years, the number of
citations of this article was greater than the corresponding number in
each of the twenty years prior to those three years.

Mark, let's go back for a moment to when Bill and I submitted the
manuscript for our 1954 joint article for possible publication in the
\textit{Journal of the American Statistical (JASA)}: Each of the
referees of the manuscript said that the manuscript should be shortened,
and the main referee said that it should be shortened by 50\%! If we had
followed the referees' instructions, our joint article definitely would
not have reached the large and wide audience that it has actually
reached and continues to reach. We decided not to follow the referees'
instructions. Instead, Bill wrote a very long, detailed letter to the
editor explaining why the manuscript should not be shortened at all---why it should be published as is. The~editor, after reading Bill's
letter, accepted the manuscript for publication as is. (By the way Mark,
the editor of \textit{JASA} at that time was Allen Wallis, who was also
at that time the first chairman of our nascent Department of
Statistics.) [Smile]

\textbf{Becker:} There is another joint article that you wrote, this one
with Ted Anderson, that also continues to be highly cited. What is that
article about?

\textbf{Goodman:} Ted and I are the coauthors of an article on
``Statistical Inference about Markov Chains'' (\textit{Ann. Math.
Stat.} 1957). This joint article is Ted's most cited article and my
second most cited article. Our joint article developed and extended the
theory and methods presented earlier by Ted in a 1954 article, and it
also presented some newer methods, which were first presented by me in a
preliminary report at the 1955 meeting of the Institute of Mathematical
Statistics. According to the ISI, our joint article has been cited
around 600 times so far. As was the case with the 1954 joint article
that Bill and I wrote, Ted's and my 1957 joint article still continues
each year to be cited by a wide range of different articles in journals
that cover a very wide range of different fields of study. In each of
the past two years, the number of citations of our 1957 joint article
was greater than the corresponding number in each of the fifteen years
prior to those two years, except for one of those fifteen years.

\textbf{Becker:} You have written many other articles on many other
topics over these many years, in addition to the joint article that you
wrote with Bill Kruskal and the joint article that you wrote with Ted
Anderson. And you have received some special recognition of this by the
Institute for Scientific Information (ISI). What was that special
recognition from the ISI?

\textbf{Goodman:} A few years ago, the ISI informed me that I have been
identified as an ``ISI Highly Cited Researcher.'' The~institute stated
that it has identified the ``250 most cited researchers in the last two
decades, for their published articles in the Mathematics category.'' For
the Mathematics category, citations of the researcher's articles
published in mathematics journals are considered, with statistics
journals included in that category.

Well, Mark, I~have been around for quite a long time, and I have written
quite a few articles---over 150 articles so far---and I have had
published four different books (each book a gathering of my articles on
a particular topic, with one of the books a gathering of Bill's and my
joint articles).

\textbf{Becker:} Leo, the eminent quantitative sociologist, Otis Dudley
Duncan, has published comments on your statistical contributions in four
different research areas that are of interest to sociologists and to
some other social and behavioral scientists. How did this come about?

\textbf{Goodman:} The~American Sociological Association (ASA) had
selected Dudley and me to share their main methodology award, the Samuel
A. Stouffer Methodology Award. When I heard this news, I~felt pleased
and honored, especially since I thought very highly of Dudley's research
work. However, Dudley had a quite different reaction to this news. He
wrote a statement, which was published in the \textit{ASA Footnotes},
that commented on my research work in four different areas of interest,
and he said that it would be a great honor to share the award with me,
but he felt strongly that I should be the sole recipient of the award,
and that the honor should not be diluted. He turned down the award.

\textbf{Becker:} What were Duncan's comments about your work in those
four areas of
interest?

\textbf{Goodman:} The~four areas were the following: (1)~social
stratification and mobility, (2) survey data analysis, (3) panel studies
data analysis, (4) latent structure analysis.

With respect to social stratification and mobility, Dudley said that my
work on methods for analyzing social mobility tables had solved a
problem that had plagued research workers in this field for at least two
decades; and, in solving this problem, my work had rendered obsolete a
substantial corpus of previous work.

With respect to survey data analysis, he said that my collection of
models for survey analysis had provided for the first time a set of
statistical methods that were adequate to the tasks posed by the
``language of social research'' hitherto associated with the Columbia
school and kindred approaches to survey analysis; and that the practiced
user of my methods could accomplish with ease everything that this
school attempted, and a great deal more. He also said that almost any
complex body of data previously analyzed by even skilled practitioners
of survey analysis, of the kind associated with the\break Columbia School and
kindred approaches, yields different conclusions from those obtained
with my
methods, and that it is easy to see after the fact how the practitioner
fell into error.

With respect to panel studies data analysis, he said that I had put
panel analysis on a sound footing for the first time, in a similar way
to what I had done with survey analysis, and, as a consequence, a
substantial body of previous misguided literature that provided
erroneous, misleading or merely useless procedures for manipulating
panel data and survey data could now be ignored.

And, with respect to latent structure analysis, Dudley said that I had
provided a substantial statistical foundation for latent structure
models, and that the methods of analysis that had been suggested
earlier, and that had been applied over the preceding 25 or 30 years by
various research workers, had not been satisfactory; but now, using the
methods introduced in the statistical foundation that I had presented,
it was possible to begin to understand correctly what is at stake.

Dudley's statement referred to work done by me in those four different
but related research areas, published in various statistics journals and
sociology journals. I~have continued to work in one or more of these
four research areas from time to time, and also in other research areas
as well, and I recently returned to all four of those research areas
when I was invited by the Editorial Committee of the \textit{Annual
Review of Sociology} to write the lead article for their 2007 volume. I~then wrote an article for the \textit{Annual Review} on ``Statistical
Magic and/or Statistical Serendipity: An Age of Progress in the Analysis
of Categorical Data.'' This article describes in simple terms some of
the major concepts of categorical data analysis (CDA) that have been
useful in the analysis of sociological data, examples of which include
data in the area of social stratification and mobility, and in many
other areas that make use of survey data and/or panel studies data, and
in the empirical study of latent types, latent variables and latent
structures. The~exposition in that article does not make use of any
mathematical formulas. Simple numerical examples, constructed for
expository purposes, are used as an aid in describing CDA concepts,
including quasi-independence, quasi-symmetry, symmetric association and
uniform association, which are concepts useful in the analysis of social
mobility tables, log-linear models useful in the analysis of survey
data, recursive models useful in the analysis of panel studies data and
latent class models useful in the analysis of latent structures.

\textbf{Becker:} You told us earlier about your two most cited articles,
the 1954 joint article with Bill Kruskal and the 1957 joint article with
Anderson. What about other works, which ones stand out in your mind as
having had a particularly significant impact?

\textbf{Goodman:} The~third most cited has been my main article
introducing log-linear models, ``Multivariate Analysis of Qualitative
Data: Interactions Among Multiple Classifications'' (\textit{JASA},
1970). It has been cited around 450 times so far.

Mark, while I was writing my 1970 \textit{JASA} article, I~also
developed a computer program in order to apply the log-linear models,
which were introduced in the article, to analyze in the article the set
of data in the multidimensional contingency table presented there. (This
computer program was also then used in some of my other articles on
log-linear models in order to analyze the different sets of data in the
multidimensional contingency tables presented in those articles; e.g.,
in \textit{Technometrics}, 1971; \textit{American Sociological}
\textit{Review}, 1972; \textit{American Journal of Sociology}, 1972,
1973; \textit{Biometrika}, 1973; \textit{JASA}, 1973.) When I was
developing the computer program, I~wanted to make it easy to use by
anyone---even someone who had no familiarity with statistics or hardly
any familiarity with the subject. One of our graduate students at that
time, Robert (Bob) Fay, helped me make the computer program as
user-friendly as was possible at that time. The~computer program was
called \textit{ECTA}: \textit{Everyman's Contingency Table Analyzer.}
(Mark, perhaps I should have called it ``Everyperson's'' or ``Everyman's
and Everywoman's'' rather than just ``Everyman's''; but at that time, I~viewed ``Everyman'' to mean also ``Everyperson'' and/or ``Everyman and
Everywoman.'') [Smile]

Copies of \textit{ECTA} were then sold at cost by our statistics
department at the University of Chicago. It is my impression that, over
the years, as many as five hundred people from many countries all over
the world purchased copies. Now most statistical computer packages
include a log-linear program.

Five years after my 1970 \textit{JASA} article appeared in print, Shelby
Haberman's monograph, \textit{The~Analysis of Frequency Data}, was
published, and it made substantial theoretical contributions to
log-linear modeling, This monograph was based on the earlier Ph.D.
thesis that Shelby wrote when he was a graduate student in our
statistics department at the University of Chicago. (I was the advisor
on his Ph.D. thesis.) And starting six years after my 1970 article
appeared in print, and continuing until now, many good textbooks,
covering log-linear analysis and other related topics in the analysis of
categorical data (frequency data, qualitative data), were published; and
many statistics departments began to introduce courses covering these
subjects in their curriculum.

Mark, returning now to your question about which of my other works
(besides the Goodman/Kruskal article and the Anderson/Goodman article)
had a significant impact, in addition to my 1970 \textit{JASA} article
on log-linear models, there are also eleven articles of mine, each of
which has been cited between 200 and 400 times so far. I~think it is
interesting, and sometimes surprising, to see which articles appear in
this group, and which do not.

\textbf{Becker:} Leo, I~will include the articles in this highly cited
group, and also some of the articles that are not in this group, as an
\hyperref[append]{Appendix} to this interview, which will be located at the end of the
interview. Here now is a question about another important set of your
articles:

You have written many articles developing modeling frameworks, and the
corresponding statistical methods, for the analysis of multidimensional
cross-classifica\-tions of counts. I~have in mind here your evolutionary
path from log-linear models to the recursive-models modeling framework,
the latent-class modeling framework, the association-models
(log-bilinear) modeling framework and the\break correspondence-analysis
modeling framework. A hallmark of these works is that they are expertly
crafted and exquisitely illustrated with insightful applications. What
has been your approach to developing these modeling frameworks?

\textbf{Goodman}: In my 2007 \textit{Annual Review of Sociology} article
on ``Statistical Magic and/or Statistical Serendipity: $\ldots,$'' which I
mentioned briefly earlier, I~discuss two different methods that I have
used to obtain the results presented in those many articles on modeling
formulations, namely, statistical magic and statistical serendipity.
[Smile/Laughter]

First, let me comment on magic: By a magical result, I~don't mean here a
result obtained by magic or by some other supernatural means, but rather
a result obtained as if by magic.

When the great Michelangelo was sculpting his colossal figure of
\textit{David}, he worked under the premise that the image of
\textit{David} was already in the block of marble that he had selected,
and his task was to release the image from the block. Now, faced with a
set of categorical data of interest, say, a multidimensional
cross-classification of counts, data analysts can work under the premise
that there is an image, or more than one image, embedded in the set of
data, and their task is to release that image, or those images, using
suitable tools.

The~tools might include the kinds of modeling frameworks developed in
some of my articles and also other kinds of tools of categorical data
analysis. The~results obtained by using these tools sometimes seem
magical---the sudden release of form formerly hidden, embedded in a
block of dense data.

Now, let me comment on serendipity: By a serendipitous result, I~don't
mean here a result obtained simply by accident or chance, but rather a
result obtained by an accidental exposure to information and an
application of the prepared mind. Perhaps serendipity, rather than
magic, better describes the way in which the modeling frameworks were
developed. [By the way, Mark, I~discuss the possible meanings of
\textit{serendipity} in ``Notes on the Etymology of \textit{Serendipity}
and Some Related Philological Observations'' (\textit{Modern Language
Notes}, 1961).]

When I first developed the general concept of quasi-independence and the
corresponding iterative procedure needed to apply this modeling
formulation in the analysis of data in a cross-classification of
interest, and, in particular, in the analysis of social mobility tables
(see, e.g., \textit{JASA}, 1968), the information to which I was exposed
in my work on this one statistical problem, and on the corresponding set
of substantive areas of interest, then led me to look at a second set of
substantive areas of interest and to develop the general concept of the
log-linear model and the corresponding iterative procedure needed to
apply this modeling formulation in the analysis of data in a
multidimensional cross-classification of interest, and, in \mbox{particular},
in the analysis of survey data (see, e.g., \textit{JASA}, 1970). And the
information to which I was exposed in my work on this statistical
problem, and on the corresponding second set of substantive areas of
interest, then led me to look at a third set of substantive areas of
interest pertaining to recursive models, and to develop the
corresponding iterative procedure needed to apply this modeling
formulation in the analysis of data in a multidimensional cross-classification table in which some variables (dimensions) are posterior
to other variables (dimensions), and, in particular, in the analysis of
panel studies data (see, e.g., \textit{Biometrika}, 1973). And the
information to which I was exposed in my work on this statistical
problem and on the preceding statistical problem, and on the
corresponding sets of substantive areas of interest, then led me to look
at a fourth set of substantive areas of interest pertaining to latent
types, latent variables and latent structures, and to develop the
corresponding iterative procedure needed to apply this modeling
formulation in the analysis of data in a multidimensional
cross-classification \mbox{table} in which some variables (dimensions) are
observable, and some of the variables (dimensions) are unobservable
(latent) (see, e.g., \textit{Biometrika}, 1974). Then the log-linear
models formulation applied to the two-way cross-classification of
interest and the corresponding iterative procedure led me to develop the
log-bilinear models (viz., the association models) formulation and the
corresponding iterative procedure needed to apply this modeling
formulation (see, e.g., \textit{JASA}, 1979). And the association models
formulation applied to the two-way cross-classification of interest
then led me to develop the correspondence-analysis model formulation
applied to the two-way and the $m$-way ($m>2$) cross-classification of
interest, and to develop further the association models formulation
applied to the $m$-way ($m>2$) cross-classification of interest, and the
corresponding iterative procedure needed to apply these modeling
formulations (see, e.g., \textit{Ann. Stat.}, 1985;
\textit{International Stat. Rev.}, 1986; \textit{JASA}, 1991).

This step-by-step movement from one statistical problem to the next
statistical problem, and from the corresponding iterative procedure
appropriate for the one statistical problem to the corresponding
iterative procedure appropriate for the next statistical problem, might
be described as step-by-step evolutionary elaboration.

\textbf{Becker:} What are some of the interesting experiences you have
had in connection with the writing of some of your articles?

\textbf{Goodman:} Mark, the following experience took place having, as
the background setting, the Cold War between the Soviet Union and the
United States: [Smile] In the late 1950s, I~wrote a number of
statistical articles pertaining to Markov chains (\textit{Ann. Math.
Stat.}, 1958a, 1958b, 1959; \textit{Biometrika}, 1958), and after those
articles were published, I~happened to come across a 1957 note by V. E.
Stepanov on Markov chains, written in Russian and published in the
Soviet journal, \textit{Teoriya Veryatnostei i ee Primeneniya}
(\textit{The~Theory of Probability and Its} \textit{Applications).}
(Stepanov was one of the Russian mathematician/probabilists who worked
in the famous Russian school of probability, founded by Kolmogorov,
Markov, Kinchin and Lyapunov.) I wasn't able to read the Russian in the
Stepanov note, but I could read the formulas, and I could see from the
formulas that the topic covered in the Russian note was very similar to
the topic that I had covered in some of my articles. So I had the
Russian note translated, and I then studied the translated version. And
I found that there was a serious error in the note, and that the main
result was incorrect. I~then wrote ``A Note on Stepanov's Test for
Markov Chains,'' showing what was incorrect in Stepanov's note and also
showing how what was incorrect could be replaced by a corresponding
correct result. I~sent it to the Editor of the Soviet journal for
submission, and received a very quick response saying that it would be
published ``in the nearest possible future.'' When the galley proof
arrived, I~noticed that my Abstract had been deleted and replaced by a
Russian Abstract in which any reference to the error in Stepanov's note
and the correction in my note had been deleted. The~Russian Abstract was
misleading. So I had my Abstract translated into Russian, and I sent it
to the Editor informing him that the Russian Abstract that was in the
galley proof needed to be replaced by my Russian Abstract. Now, Mark,
what do you think the Editor did about this?

Well, the Editor didn't respond to my request, and he didn't pay any
attention to my Russian Abstract. When my note was published in the
Soviet journal, I~then saw that the note included the misleading Russian
Abstract that was in the galley proof, and it also included something
else that really surprised me. In addition to the misleading Russian
Abstract, it also included a second Abstract, my original Abstract
(which had been written in English) printed in English in my note!

Here is my conjecture as to why the Editor included the two Abstracts in
the published note: I think that he included the misleading Russian
Abstract because he didn't want any nontechnical Russian readers to know
that an American statistician was critical of work done by a Russian
mathematician/probabilist; and he included my Abstract in English
because he wanted technical readers, who might be interested in the
topic, to know as much as possible about the topic.

Here now is another interesting experience that I had: This experience
takes place having, as the background setting, the early 19th century
French philosophers. [Smile] In 1819, the Marquis de Laplace, in
\textit{A~Philosophical Essay} \textit{on Probabilities} (title
translated from French), discussed the attempts that had been made still
earlier to explain the excess of the birth of boys over those of girls
by the general desire of fathers to have a son. Laplace's results
suggested that the sex ratio at birth of boys to girls will be
unaffected by this general desire. In the early 1950s, the statistician
Herb Robbins arrived at a similar conclusion. However, in the early
1960s, I~found, in the sociological literature, first, an article by a
sociologist who suggested that, for the particular group of families he
had studied, the prevalence of the desire for male offspring on the part
of parents, together with their knowledge of methods of birth-control,
appeared to be significant in relation to the high sex-ratio at birth of
boys to girls; and, second, an article by another sociologist who proved
that the sex-ratio at birth of boys to girls will be either decreased or
unaffected by the preference for male offspring, if certain assumptions
can be made concerning the way in which this preference affects the
parents' decisions as to whether or not to have another child.

Taking all this into account, I~was able to reconcile the different
conclusions obtained by these different authors in an article that I
wrote on ``Some Possible Effects of Birth Control on the Human Sex
Ratio,'' in which I presented a general framework for the study of these
possible effects that included as special cases each of the possible
assumptions that might be made in this context, and I introduced
formulas that show under which possible assumptions the sex ratio would
be unaffected, under which assumptions it would be decreased, and under
which assumptions it would be increased. The~article was published in
the \textit{Ann.} \textit{Human Genetics (London)} in 1961, and it was
reprinted in 1966 in a volume on \textit{Readings in Mathematical Social
Science.}

Next is another interesting experience that I had: It took place having,
as the background setting, James Bond, Agent 007. [Smile] In the early
1950s, I~came across an article that described how the Allies in World
War II analyzed serial numbers obtained from captured German equipment
in order to obtain estimates of German war production and capacity.
Within the limits of its capabilities, this method of analysis turned
out to be superior to more abstract methods of intelligence. After
reading that article, I~thought that I would try, for the fun of it, to
see if I could improve on the method of analysis of the serial numbers
that was described in that article. Well, it happened to turn out that I
could improve on it. With my method of analysis of the serial numbers, I~was able to introduce a simple estimator that was the most efficient
unbiased estimator of the corresponding total number of pieces of
equipment in the population of pieces of equipment from which the
observed serial numbers came. I~was also able to show that the method of
estimation described in the article on the World War II method of
analysis provided an estimator that was unbiased, and that the
efficiency of that estimator was relatively high for large or moderate
sized samples of serial numbers. I~wrote up these results, together with
some other statistical results on this subject, in ``Serial Number
Analysis'' (\textit{JASA}, 1952). About five or so years after the
publication of that article, although he should not have told me about
this, someone whom I knew told me that there was a group of people in
the government in Washington, D.C., who were using what they called the
``Goodman method,'' making use of the results in that article. I~was
surprised by this news, although perhaps I should not have been.

Now let me tell you about just one more interesting experience, which I
had just last year: [Smile] The~U.S. Court of Appeals for the Ninth
Circuit is the largest of the thirteen U.S. circuit courts of appeals.
Over the past thirty years or so, many different legislative proposals
to split the Ninth Circuit have been introduced in the U.S. Congress (in
both the Senate and in the House of Representatives), and each of these
proposals has failed to become law. There has been no consensus within
Congress. The~debate on this subject was revisited a year ago in The~Los
Angeles Times (Opinion, July 11, 2007), this time with a statistical
argument purporting to conclude that the Ninth Circuit Court should be
split. Then the Circuit Executive of the Ninth Circuit Court contacted
me to inquire whether I thought that there could be a rejoinder to the
statistical argument in the LA Times. I~replied that there could be, and
I then wrote a rejoinder in the form of an OpEd statement that I
submitted for publication in the LA Times. It was rejected. I~then
submitted it in turn to two law newspapers, and it was also rejected by
each of them. (The~newspapers' rejections were explained this way: The~Republicans are no longer the majority party in Congress, and it was
primarily Republicans in Congress who, over the past thirty years or so,
had tried to split the Ninth Circuit. So there is less interest in this
subject now that the Democrats are the majority party.) I then wrote an
article, ``To Split or Not To Split the U.S. Ninth Circuit Court of
Appeals: A Simple Statistical Argument, Counterargument, and Critique,''
which has now been published in the \textit{Journal of}
\textit{Statistical Planning and Inference} (\textit{JSPI}, 2008).

\textbf{Becker:} Your 1974 article in \textit{Biometrika},\break ``Exploratory
latent structure analysis using both identifiable and unidentifiable
models,'' is one of my personal favorites. It brought clarity to an area
of data analysis and modeling that was cluttered at the time, and in a
very straightforward way you used what we know today as the EM-algorithm
to both provide a computational devise and theoretical insights. You
must be pleased to have both ``anticipated'' the EM-algorithm and to
have exploited it for theoretical gain.

\textbf{Goodman:} Mark, as you know, the first article on the
``EM-algorithm'' was written by Art Dempster, Nan Laird and Don Rubin,
and was published in 1977 in the \textit{J. Roy. Stat. Soc.} And, as you
said just now, my article was published in 1974 in \textit{Biometrika.}
When the data in a two-way or in an $m$-way ($m>2$) cross-classification are
analyzed using a latent-class model, the ``EM-algorithm'' described in
the 1977 article is essentially the same as the iterative procedure
introduced earlier in my 1974 article. And so, with the 1974 article
iterative procedure available, the 1997 article ``EM-algorithm'' isn't
needed to analyze a two-way or an $m$-way ($m>2$) cross-classification using
a latent-class model. The~1997 article ``EM-algorithm'' is useful in
other contexts.

I am pleased that, with my 1974 \textit{Biometrika} article and the 1997
\textit{J. Roy. Stat. Soc.} article available, the appropriate
computational device is now widely used, and the theoretical insights
are now widely understood, in the study of latent structures. (See, for
example, my 2002 article, ``Latent Class Analysis: The~Empirical Study
of Latent Types, Latent Variables, and Latent Structures,'' which is the
lead article in \textit{Applied Latent Class Analysis}, edited by
Jacques Hagenaars and Allan McCutcheon.)

\textbf{Becker:} You were a member of the Population Research Center at
the University of Chicago, and you've published in demography. I~know
that you wrote one article with (the eminent demographer) Nathan Keyfitz
who was also at Chicago at some time. Was he there when you and he wrote
that joint article? And was the work that you did in demography inspired
by him in any ways?

\textbf{Goodman:} Nathan Keyfitz was at the University of Chicago from
1963 to 1968. The~Goodman/\break Keyfitz/Pullum article on ``Family Formation
and the Frequency of Various Kinship Relationships'' was published in
\textit{Theoretical Population Biology} in 1974. Nathan and I started
working on that article during the time period when he and I were
colleagues at the University of Chicago. The~third coauthor of that
article, Tom Pullum, had been a graduate student in the sociology
department at the University of Chicago during a part of the time when
Keyfitz was there, and he received his Ph.D. degree in 1971. (I was the
advisor on his Ph.D. thesis.) In 1974, when our joint article was
published, Nathan was a professor at Harvard, and Tom was an assistant
professor there.

Nathan has had a very interesting career, and I would like now to tell
you about it. He graduated with a degree in mathematics from McGill
University in Montreal in 1934, and in 1936 he began working for the
Dominion Bureau of Statistics, the precursor to Statistics Canada, in
Ottawa, as a research statistician and later as a mathematical and
senior statistical advisor. He analyzed Canadian census schedules and
census results, and he prepared statistical surveys that examined
various characteristics of the Canadian population. He remained with the
bureau for the next 23 years. In 1952 he received a fellowship to attend
the University of Chicago, and he graduated with a Ph.D. in Sociology.
(I was one of the examiners on his oral exam, and I can attest to the
fact that he did very well indeed.) In 1963, at the age of 50, he was
invited to join the Sociology Department faculty at the University of
Chicago, and he accepted the invitation. At that point, as far as I
know, he had not expressed any special interest in the field of
mathematical demography nor in the application of mathematical tools and
computer technology to the analysis of demographic data.

After Nathan's arrival in Chicago, in one of our first conversations, I~happened to mention to him that I had written an article in mathematical
demography ten years earlier on the ``Population\break Growth of the Sexes''
(\textit{Biometrics}, 1953), and that I thought that much more work
could be done on this topic and in other areas of mathematical
demography as well. He and I then began to stimulate each other to do
research in this field. He then wrote many very good articles and some
very good books in this field, and I also wrote some articles in this
field, and in some other related fields too. He and I became good
friends. He became a very important leader in the field of mathematical
demography and a pioneer in the application of mathematical tools and
computer technology to the analysis of demographic data.

After Nathan taught at the University of Chicago, he then taught at the
University of California at Berkeley for four years (1968--1971), and
then he taught at Harvard for twelve years (1972--1983). He then spent
ten years (1984--1993) as the Project Leader of the Population Project at
the International Institute of Applied Systems Analysis (IIASA) in
Austria, before returning to Cambridge. Between 1973 and 1993, seven
different universities awarded him honorary doctorates, and in 1997 he
was named the International Union for the Scientific Study of Population
(IUSSP) Laureate. Isn't it interesting, Mark, that all this began after
Nathan's arrival at the University of Chicago as a faculty member at the
age of 50? He is now over 95 years of age, and in good shape considering
his age.

I wrote twelve articles in demography and related fields, one of which
was the 1974 joint article with Nathan and Tom. The~twelve articles
appear in seven different journals, and they cover a wide range of
topics. Two of the articles were published before Nathan's arrival on
the faculty at the University of Chicago, and seven of the articles were
published during the six year period when he and I were colleagues
there.

\textbf{Becker:} Leo, I~will include a reference to your articles in
demography and related fields in the \hyperref[append]{Appendix}. Now let us move from
demography and related subjects to economics.

W. Allen Wallis, the first chairman of the Statistics Department at the
University of Chicago, was both a statistician and an economist. You
wrote an article related to a joint article that Wallis wrote with
Geoffrey Moore, an expert in economic statistics and business cycle
research, and you also wrote some other articles of special interest to
economists. How did this come about?

\textbf{Goodman:} One day, sometime in the second half of the 1950s,
Allen happened to be telling me about this joint article that he had
written with Geoffrey Moore about fifteen years earlier, which
introduced a time-series significance test concerning the correlation
between the movements in two different time-series, based on the signs
of the first differences in each of the time-series. He told me that, in
the joint article, he and Moore described the conditions under which
their test would be valid, but they were aware of the fact that those
conditions were actually not realistic conditions---those conditions
wouldn't be satisfied by real economic time-series. He and Moore noted
in their article that more research was needed in order to find out how
their test would need to be modified so that the modified test would be
valid under more realistic conditions. Allen then told me that, even
though about fifteen years had now gone by since the publication of the
joint article, no one yet had successfully solved this problem.

And so I then worked on this problem for a while, and I happened to
solve the problem. I~then invited a good friend of mine, an economist,
Yehuda Grunfeld, to join me in writing a joint article that would
introduce an appropriately modified time-series significance test
concerning the correlation between the movements between two different
time-series (a new test valid under realistic conditions), and that
would apply this test to some economic time-series of interest. We then
wrote our joint article on ``Some Nonparametric Tests for Comovements
Between Time-Series'' (\textit{JASA}, 1961).

About eight months before our joint article appeared in print, a
terrible tragedy occurred: Yehuda died in a drowning accident at the age
of 30. At the time of his death, he was Lecturer in Economics and
Statistics at the Hebrew University of Jerusalem, and earlier he had
been a graduate student and then an Assistant Professor in the Economics
Department at the University of Chicago. In memory of Yehuda, a volume
was published, \textit{Measurement in Economics: Studies in}
\textit{Mathematical Economics and Econometrics}, in 1963. Authors
contributing to this volume included the Nobel Laureate Milton Friedman
and other top economists and econometricians. A section on Econometric
Methodology was included in the volume, and I wrote on ``Tests Based on
the Movements in and the Comovements between $m$-Dependent Time-Series,''
which was included as the first article in that section.

The~results presented in my article on movements in and comovements
between time-series were a further development and extension of results
presented in the earlier joint article that Yehuda and I wrote on
comovements, which could be viewed as a further development and
extension of results presented in the Wallis/Moore joint article.

Mark, I~would like now to tell you about Allen Wallis. During World War
II, Allen, in his early 30s, served as director of research in a
Statistical Research Group of the U.S. Office of Scientific Research,
and he recruited a stellar group of young statisticians, mathematicians
and economists who contributed significantly to the war effort in many
ways. Before the war, Allen taught very briefly at Yale, Columbia and
Stanford universities; and after the war, he went back very briefly to
Stanford, and then he became a faculty member at the University of
Chicago. As I said earlier, Allen was responsible for establishing our
statistics department at the University of Chicago, and he was the first
chairman of the department. He then became the dean of the University of
Chicago Graduate School of Business, and then he moved to the University
of Rochester as president and then chancellor, and after he retired
there, he served as Under Secretary of State for Economic Affairs in the
Reagan administration. He also served as an economic advisor to four
U.S. presidents, Eisenhower, Nixon, Ford and Reagan.

Well, Mark, let's return now to your question about the work that I did
of interest to economists. In addition to my article on movements in and
comovements between time series, and my joint article with Yehuda, I~also wrote a joint article with another economist, Harry Markowitz, who
later was awarded a Nobel Prize in Economics. (Strictly speaking, the
prize in Economics is actually the ``Bank of Sweden Prize in Economic
Sciences in Memory of Alfred Nobel.'') The~prize was for Harry's
pioneering work in financial economics, in modern portfolio theory, and
in studying the effects of asset risk, correlation and diversification
on expected investment portfolio returns.

In the early 1950s, Harry was a graduate student in the Economics
Department at the University of Chicago, and he was also a Research
Fellow at the Cowles Commission for Research in Economics,\break  which was
affiliated with the university and with the Economics Department. At the
Cowles Commission at that time, there was great interest in recent work
by the economist Ken Arrow, who had earlier been a Research Associate at
the Cowles Commission and an Assistant Professor in the Economics
Department at the university. (Arrow was also awarded a Nobel Prize in
Economics for his pioneering contributions to general economic
equilibrium theory and welfare theory.)

The~work by Arrow that was of great interest at the Cowles Commission in
the early 1950s was his research on \textit{Social Choice and Individual
Values}, in which he described five apparently reasonable properties
that any voting system or other ``social \mbox{welfare} function'' should have,
and he demonstrated mathematically that no voting system (or other
social \mbox{welfare} function) could possibly have all of these properties.
Harry and I, in our joint article, ``Social Welfare Functions Based on
Individual Rankings'' (\textit{Am. J. Soc.}, 1952), demonstrate that one
of Arrow's required properties is questionable, and, if this property is
modified, then many voting systems become acceptable. The~joint article
also considers which of the many voting systems, considered acceptable
by us, seem most reasonable.

\textbf{Becker:} You mentioned the Cowles Commission for Research in
Economics. What was your relationship with the Commission, and with some
of the other members of the Commission?

\textbf{Goodman:} The~Cowles Commission was founded to advance the
scientific study and development of economic theory in its relation to
mathematics and statistics, and it played a major role in promoting the
use of mathematics and statistics in economics. Innovative and seminal
work in mathematical economics and econometrics took place at the
Commission in the years 1943--1955, years in which the research directors
were first Jacob Marschak and then Tjalling Koopmans. The~research
output at the Commission over that period of time was extraordinary.
Nine economists who were at the Cowles Commission at some time during
the 1943--1955 period were later awarded Nobel Prizes, and I am quite
sure that there would have been ten such Nobel Laureates if Marschak had
lived a little bit longer. The~nine were Ken Arrow, Tjalling Koopmans,
Herbert Simon, Lawrence Klein, Gerard Debreu, Franco Modigliani, Trygve
Haavelmo, Harry Markowitz and Leonid Hurwicz.

%f11 ###
\begin{figure*}

\includegraphics{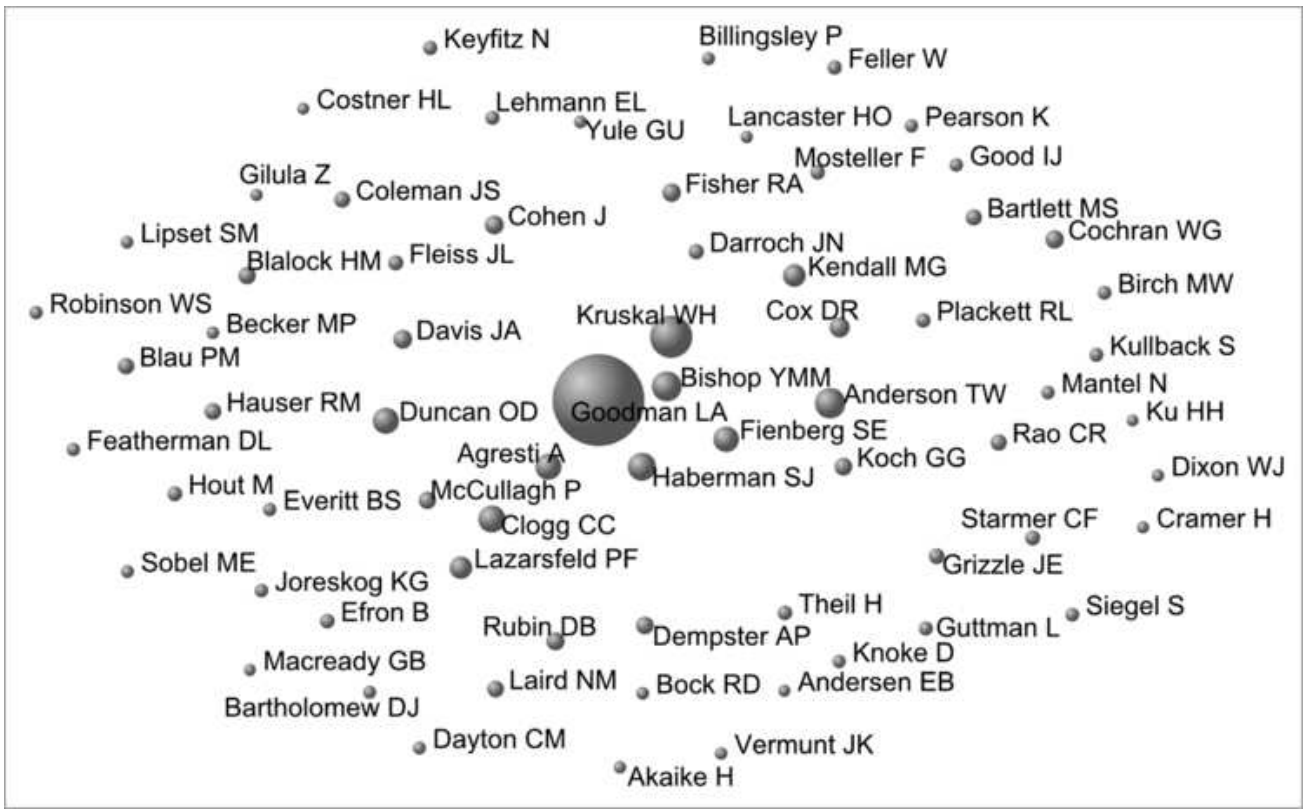}

\caption{Map of authors co-cited with Leo Goodman in at least 100
articles.}
\end{figure*}

Jacob (Yascha) Marschak and I were good friends, and he would from time
to time ask me statistical questions that he needed to deal with in his
economics research, and from time to time I knew the answers to his
questions or I could figure out the answers. Some other economists at
the Cowles Commission would also ask me statistical questions from time
to time. In 1955, the Commission moved from the University of Chicago to
Yale where it was renamed the Cowles Foundation for Research in
Economics. Yale at that time did not have a statistics department, and
there weren't any statisticians at the university who could help Cowles
Foundation members with statistical questions that came up in their
economics research.

In 1960, Kingman Brewster was selected by the Yale President, A. Whitney
Griswold, to be the\break Provost at Yale, and the Cowles Foundation then
asked Brewster to establish a statistics department there. The~Foundation also invited me to give a talk at their Economics Colloquium.
When I came to Yale to give \mbox{the talk}, I~was introduced to Brewster, and
he came to the \mbox{talk. In} the question and answer period following the
talk, Brewster asked some excellent questions. I~was very impressed. He
then invited me to meet with him for lunch at the Harvard Club in New
York. (I was at that time a visiting professor at Columbia University in
New York, on leave from the University of Chicago.)

At our first lunch meeting, we considered the possibility of
establishing a statistics department at Yale, and it seemed pretty clear
to me that Brewster was very uncertain about moving ahead with this. He
was the Provost, and he was being groomed to become possibly the next
President of Yale. If he, as the Provost at Yale, were to establish a
new department in some field, it would, of course, need to be the best
department, or at least among the very best departments, in that field
in the country. At another lunch meeting, he was telling me at one point
in our conversation that, when he was an undergraduate at Yale, in order
to be considered a Yale man---a Yale educated man---there was a
special course in philosophy that one would need to complete, taught by
a very special professor (whose name I have now forgotten). And I
responded as follows: ``Well, that was in your day at Yale. But now, in
the second half of the twentieth century, in order to become an educated
person, a Yale man could use a good course in statistics, in part, in
order to help him to avoid being misled by statements read in newspapers
or heard on radio or television, or more generally communicated by any
medium.'' After that, Brewster seemed no longer to be uncertain about
moving ahead with the possibility of establishing a statistics
department, and we began to discuss in detail what was needed in order
for this to happen. At our next lunch meeting, Brewster offered me the
job of building the department, and I thanked him for the offer. I~thought about the offer for a few days, and then told Brewster that I
was turning the offer down. He then asked me whom would I recommend for
the job, and I recommended Frank Anscombe who was at Princeton at that
time. Brewster then offered Anscombe the job, and Anscombe accepted the
offer.

\textbf{Becker:} In this conversation, we have talked to some extent
about your working relationships with some statisticians and social
scientists, sociologists, demographers and economists, and about some
joint work that you have done with some of these people. You had
mentioned to me in some earlier conversation that there was a way of
showing pictorially who were some of the people in these fields whose
works were related to your works and/or who were some of the people in
these fields to whose works your works were related. I~would be
interested to see this pictorial representation.

\textbf{Goodman:} Here it is---a map of authors who are co-cited with
me in at least 100 articles. The~information that was used to create
this map was obtained from the Institute for Scientific Information
(ISI) Web of Science, based on 5678 articles that cite articles written
by me. The~creator of this map was Olle Perrson, Professor of Sociology,
at Umea University in Umea, Sweden.

The~size of the ball associated with each author on the map is
proportional to the number of citations of that author's articles in the
5678 articles; so we see here that the largest balls were for W.~H.
Kruskal, T.~W. Anderson, Y.~M.~M. Bishop, S.~J. Haberman, C.~C. Clogg, O.~D. Duncan,
A.~Agresti, S.~E. Fienberg, M.~G. Kendall; and the next largest balls were for
P.~F.
Lazarsfeld, D.~R. Cox, J.~Cohen, J.~A. Davis, R.~A. Fisher, W.~G. Cochran, D.~B. Rubin,
G.~G. Koch, A.~P. Dempster; and the next largest balls were for H.~M. Blalock,
P.~McCullagh, N.~M. Laird, R.~M. Hauser, P.~M. Blau, C.~R. Rao, J.~S. Coleman and
M.~S.
Bartlett. The~place where an author appears on the map is based on the
number of co-citations of the author with me and also on the number of
co-citations of the author with each of the other authors on the map. It
turns out that the authors on the right side of the map are mainly
statisticians, and there also are some mathematician/probabilists and
some others there; and the authors on the left side of the map are
mainly sociologists, and there also are some statisticians and some
others there.

%f12 ###
\begin{figure*}[b]
\begin{tabular}{cc}

\includegraphics{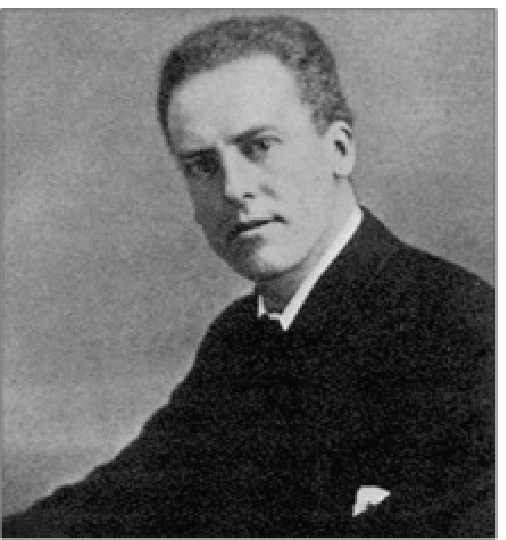}
&\includegraphics{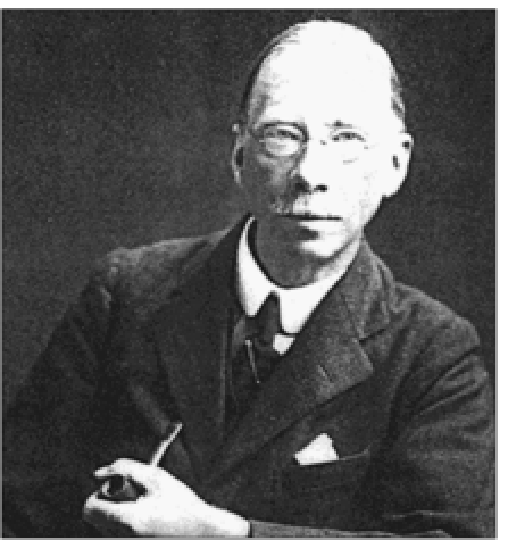}\\
Karl Pearson& G. Udny Yule\\[6pt]

\includegraphics{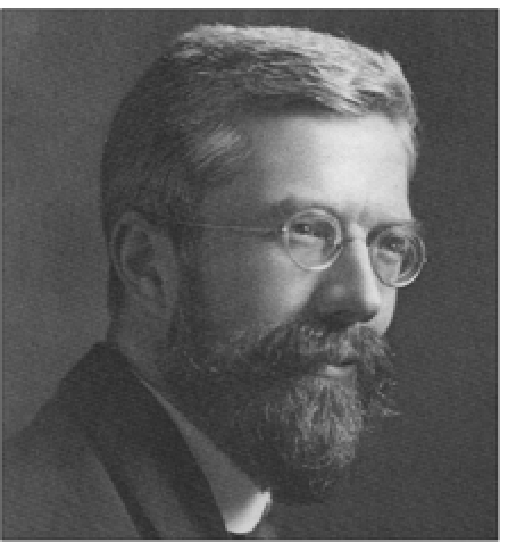}
&\includegraphics{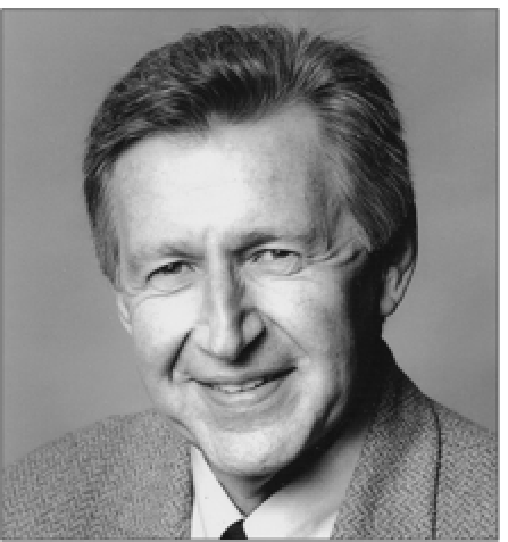}\\
Ronald A. Fisher& Leo Goodman
\end{tabular}
\caption{Four leading figures in the development of categorical data
analysis.
Source: Alan Agresti, ``Categorical Data Analysis'' 2nd~edition, 2002; and
``An Introduction to Categorical Data Analysis'' 1st and 2nd edition, 1996
and 2007.}
\end{figure*}

\textbf{Becker:} Leo, I~can see how this kind of map could be useful in
many different contexts. Thanks for bringing this to my attention. Now
there is just one more topic that I feel I need to bring up with you
before our conversation comes to an end. I~know that you are a cancer
survivor, now more than 30 years. How did your battle with and victory
over cancer influence your work?

\textbf{Goodman:} After the course of treatment for my cancer was
completed, I~then wrote my main article introducing and developing
association models: ``Simple models for the Analysis of Association in
Cross Classifications Having Ordered Categories'' (\textit{JASA}, 1979). I~view the contents of this article as a big step forward in categorical
data analysis (CDA). Somehow, having my mind completely focused, during
the course of the cancer treatment, on doing whatever I could to
increase the chances that I might become a cancer survivor, this
focusing of my mind then helped me later to clear my mind, after the
course of treatment was completed, and to take a big step forward when I
was able to return to statistical work. [In Alan Agresti's first and
second edition of his book on \textit{Categorical Data Analysis}
(Agresti, 1990, p. 505, and 2002, pp. 631), my 1979 \textit{JASA}
article is included in his list of 25 articles that convey a sense of
how CDA methodology had evolved during the twentieth century. (By the
way Mark, Agresti also lists, in the second edition of his book on
\textit{Categorical Data Analysis}, Karl Pearson, G. Udney Yule, Ronald
A. Fisher and Leo Goodman, as ``Four leading figures in the development
of categorical data analysis.'')] My 1979 article was included as the
core article in my 1984 book on \textit{The~Analysis of Cross-Classified
Data Having} \textit{Ordered Categories}, and I also extended the work
presented in the 1979 article in some of my later work---\textit{JASA},
1981a, 1981b, 1991, 1996; \textit{Ann. Stat.}, 1985;
\textit{International Stat. Rev.}, 1986; \textit{Amer. J.}
\textit{Soc.}, 1987.

Mark, with respect to the cancer, here's an interesting experience that
I had. This experience leads me to say sometimes that \textit{it
was statistics that saved my life}. [Smile] Here's
what happened: After the surgical removal of the cancer in New York,
there was a disagreement between my New York oncologist and a group of
oncologists in Chicago as to what should be done next. The~New York
oncologist said that, for the particular kind of cancer that I have, a
course of chemotherapy and immunotherapy should be administered at once;
and the Chicago group of oncologists said that, for the particular kind
of cancer that I have, a course of radiation should be administered at
once, and that chemotherapy and immunotherapy should not be
administered. The~Chicago group of oncologists gave me a number of
articles to read on this subject. These articles had been published in
various British medical journals, and the abstract in each of the
articles stated that, with this kind of cancer, radiation was
recommended. I~then studied carefully the text of each of these
articles, and it seemed to me that the detailed medical and statistical
evidence presented in the articles themselves did not warrant the
recommendation presented in the abstracts. So I returned to the Chicago
group to ask them some questions about the articles, and their responses
to the questions left me with the impression that they had read the
abstracts but they had not studied carefully the articles themselves. I~then decided to follow the New York oncologist's regimen.

It turned out that the New York oncologist's regimen was somewhat
similar to what was done at that time in France for this kind of cancer,
and the Chicago group's regimen was similar to what was done at that
time in Britain. A few years after I had completed the New York
oncologist's regimen, it turned out that an international medical
conference was held on ``Cancers of the Mid-East,'' and comparisons were
made there, for those who had the kind of cancer that I had, between
mortality statistics for those receiving the British regimen in Britain
and those receiving the French regimen in France. For the British
patients, the death rate was really terrible, whereas the death rate for
the French patients was not as bad. Mark, imagine what might have
happened if I had just read the abstracts in the various British medical
journals, and had not bothered to study carefully the detailed medical
and statistical evidence presented in the articles themselves? [Smile]

\textbf{Becker:} Leo, thank you for being so generous with your time and
reflections on your almost six decades statistical career. And what a
career it has been---the experiences and relationships that you have had
have been nothing short of amazing.

\begin{appendix}\label{append}

\section*{Appendix A}
As was noted earlier in this conversation, there
are eleven articles each cited between 200 and 400 times so far. These
articles will be described here (but not necessarily in the order
pertaining to the number of citations of each article): (1) the R. A.
Fisher Memorial Lecture, ``The~Analysis of Cross-Classified Data:
Independence, Quasi-Independence, and Interactions in Contingency Tables
With or Without Missing Entries'' (\textit{JASA}, 1968); (2--3) two
articles introducing new methods for the analysis of latent structures,
``Exploratory Latent Structure Analysis Using Both Identifiable and
Unidentifiable Models'' (\textit{Biometrika}, 1974), and ``The~Analysis
of Systems of Qualitative Variables When Some of the Variables Are
Unobservable: A Modified Latent Structure Approach,'' \textit{Amer. J.
Soc.} (\textit{AJS}, 1974); (4) an article introducing association models,
``Simple Models for the Analysis of Association in Cross-Classifications
Having Ordered Categories'' (\textit{JASA}, 1979); (5) an article
introducing various procedures for using log-linear models to fit
contingency-table data, ``Analysis of Multidimensional Contingency
Tables: Stepwise Procedures and Direct Estimation Methods for Building
Models for Multiple Classifications'' (\textit{Technometrics}, 1971);
(6--7) two articles introducing multiplicative models to analyze
categorical data, ``A General Model for the Analysis of \mbox{Surveys}''
(\textit{AJS}, 1972), and ``A Modified Multiple Regression Approach to
the Analysis of Dichotomous Variables,'' \textit{Amer. Soc.}
\textit{Rev. \textup{(}ASR}, 1972); (8--9) the second and third joint articles
with Bill Kruskal, ``Measures of Association for Cross Classifications
II: Further Discussion and References'' (\textit{JASA}, 1959), and
``Measures of Association for Cross Classifications III: Approximate
Sampling Theory'' (\textit{JASA}, 1963); (10) an article on some methods
for dealing with the ecological-correlation problem, ``Some Alternatives
to Ecological Correlation'' (\textit{AJS}, 1959); (11) an article
introducing exact formulas for the variance of products, and formulas
for estimating this variance, ``On the Exact Variance of Products''
(\textit{JASA}, 1960).

There are twelve articles each of which has been cited between 100 and
200 times so far. These articles will be described next (but not
necessarily in the order pertaining to the number of citations of each
article): (1) the Henry L. Rietz Memorial Lecture, ``The~Analysis of
Cross Classified Data Having Ordered and/or Unordered Categories:
Association Models. Correlation Models, and Asymmetry Models for
Contingency Tables With or Without Missing Entries'' (\textit{Ann.
Stat.}, 1985); (2) an article introducing correspondence analysis
models, ``Some Useful Extensions of the Usual Correspondence Analysis
Approach and the Usual Log-Linear Models Approach in the Analysis of
Contingency Tables'' (\textit{International} \textit{Stat. Rev.}, 1986);
(3--4) two articles introducing recursive models for panel analysis,
``The Analysis of Multidimensional Contingency Tables When Some
Variables Are Posterior to Others: A Modified Path Analysis Approach''
(\textit{Biometrika},\break 1973), and ``Causal Analysis of Data from Panel
Studies and Other Kinds of Surveys,'' \textit{Amer. J. Soc. \textup{(}AJS},
1973); (5) an article introducing methods for the analysis of data
obtained by snowball sampling, ``Snowball Sampling'' (\textit{Ann. Math.
Stat.}, 1961); (6) a joint article with Clifford Clogg, introducing
latent structure models for analyzing simultaneously more than one
multidimensional contingency table, ``Latent Structure Analysis of a Set
of Multidimensional Contingency Tables'' (\textit{JASA}, 1984); (7) an
article introducing additional methods for analyzing mobility tables,
``How to Ransack Social Mobility Tables and Other Kinds of
Cross-Classification \mbox{Tables}'' (\textit{AJS}, 1973); (8) an article
introducing new methods for analyzing scales, ``A New Model for Scaling
Response Patterns: An Application of the Quasi-Independence Concept''
(\textit{JASA}, 1975); (9) the fourth joint article with Bill Kruskal,
``Measures of Association for Cross Classification IV: Simplification of
\mbox{Asymptotic} Variances'' (\textit{JASA}, 1972); (10) an article describing
the relationship between RC association models and canonical
correlation, ``Association Models and Canonical Correlation in the
Analysis of Cross Classifications Having Ordered Categories''
(\textit{JASA}, 1981); (11--12) two articles on simultaneous confidence
intervals, ``Simultaneous Confidence Intervals for Contrasts Among
Multinomial Populations'' (\textit{Ann. Math. Stat.}, 1964), and ``On
Simultaneous Confidence Intervals for Multinomial Proportions''
(\textit{Technometrics}, 1965).

There are fifteen articles each of which has been cited between 50 and
100 times so far. Ten of these articles will be described next (but not
necessarily in the order pertaining to the number of citations of each
article): (1) an article, based on an invited lecture, presented at the
invitation of the Amer. Stat. Assoc. Social Statistics Section, on
correspondence analysis models and related topics, with comments by J.~P.
Benzecri, the founder of and major figure in the ``French school of data
analysis'' (the school of correspondence analysis), and also comments by
D.~R. Cox, C.~R. Rao, S.~J. Haberman. H. Caussinus, C.~C. Clogg, and by
three others, and a rejoinder by me, ``Measures, Models, and Graphical
Displays in the Analysis of Cross-Classified Data'' (\textit{JASA},
1991); (2) a joint article with Nathan Keyfitz and Tom Pullum on kinship
relationships, ``Family Formation and the Frequency of Various Kinship
Relationships'' (\textit{Theoretical Population} \textit{Biology},
1974); (3) an article in mathematical demography, ``Population Growth of
the Sexes'' (\textit{Biometrics}, 1953); (4) an article introducing
multiplicative models for mobility table analysis, ``Multiplicative
Models for the Analysis of Occupational Mobility Tables and Other Kinds
of Cross-Classification Tables'' (\textit{AJS},1979); (5)~an article on
the relationship \mbox{between} the RC association models and the bivariate
normal distribution, ``Association Models and the Bivariate Normal for
Contingency Tables with Ordered Categories'' (\textit{Biometrika},
1981); (6) an article on additional methods for analyzing
multidimensional contingency tables, ``Partitioning of Chi-Square,
Analysis of Marginal Contingency Tables, and Estimation of Expected
Frequencies in Multidimensional Contingency Tables'' (\textit{JASA},
1971); (7) an article on the analysis of certain kinds of panel data
using the mover-stayer model,\ a~generalization of the Markov chain
model, ``Statistical Methods for the Mover-Stayer Model''
(\textit{JASA}, 1961); (8) an article presenting explicit formulas for
analyzing three-factor interaction in contingency tables, ``Simple
Methods for Analyzing Three-Factor Interaction in Contingency Tables''
(\textit{JASA}, 1964); (9) an article on simultaneous confidence-limits,
``Simultaneous Confidence-Limits for Cross-Product Ratios in Contingency
Tables'' (\textit{J. Roy. Stat. Soc.}, \textit{Ser. B}, 1964); (10) an
article developing further some statistical methods presented in an
earlier article by Ted Anderson and in the earlier joint
Anderson/Goodman article, ``Statistical Methods for Analyzing Processes
of Change'' (\textit{AJS}, 1962).

\section*{Appendix B}
Twelve articles in demography and related fields:
\textit{Biometrics}, 1953, 1969; \textit{Ann. Human Genetics (London)},
1961, 1963; \textit{Demography}, 1967, 1968; \textit{J.} \textit{Roy.
Stat.} \textit{Soc., Ser. A}, 1967; \textit{Biometrika}, 1967, 1968;
\textit{Ann. Math. Stat.}, 1968; \textit{Theoretical Population
Biology}, 1971, 1974 (the joint Goodman/Keyfitz/Pullum article).

\section*{Appendix C}
Photo Credits: The~photos of Lipman Bers, Alonzo
Church, and Emil Artin are based on photographs by Paul R. Halmos, from
\textit{I Have A Photographic Memory}, copyright 1987 American
Mathematical Society, with permission from the AMS. The~photo of Salomon
Bochner is based on a photograph from the \textit{Collected}
\textit{Papers of Salomon Bochner, Part 1}, R. C. Gunning, ed., copyright
1992 American Mathematical Society, with permission from the AMS. The~photo of Bill Kruskal is based on a photograph from the 2005 \textit{IMS
Bulletin}, with permission from the Institute of Mathematical
Statistics. The~photo of John Tukey is based on a photograph by
Elizabeth Menzies, Princeton photographer. The~photos of Charles
Loewner, Sam Wilks, and Solomon Lefschetz are based on photographs from
the archives of the University of St. Andrews, Scotland.
\end{appendix}

\vspace*{-2pt}
\end{document}